\documentclass[journal=jpcafh]{achemso}
\usepackage{multirow}
\usepackage{graphicx}
\usepackage{hyperref}
\usepackage{chemformula}
\usepackage[sort&compress,numbers,super]{natbib}
\usepackage[T1]{fontenc} 
\usepackage[utf8]{inputenc}
\usepackage{multirow}
\usepackage{amsmath}
\usepackage{amssymb}
\usepackage{subcaption}
\usepackage{caption}
\usepackage{lscape}
\usepackage[table]{colortbl}
\graphicspath{{./Figures_new/}}

\SectionNumbersOn

\author{Xiaotong Liu}
\affiliation{Beijing Advanced Innovation Center for Materials Genome Engineering, Beijing Information Science and Technology University, Beijing 100101, P.~R.~China}
\alsoaffiliation{School of Computer, Beijing Information Science and Technology University, Beijing 100101, P.~R.~China}
\author{Pierre-Paul De Breuck}
\affiliation{UCLouvain, Chemin des Étoiles 8, Louvain-la-Neuve 1348, Belgium}
\author{Linghui Wang}
\affiliation{School of Computer, Beijing Information Science and Technology University, Beijing 100101, P.~R.~China}

\author{Gian-Marco Rignanese}
\affiliation{UCLouvain, Chemin des Étoiles 8, Louvain-la-Neuve 1348, Belgium}
\email{gian-marco.rignanese@uclouvain.be}

\title{A simple denoising approach to exploit multi-fidelity data for machine learning materials properties}

\begin{document}

\maketitle

\begin{abstract}
Machine-learning models have recently encountered enormous success for predicting the properties of materials.
These are often trained based on data that present various levels of accuracy, with typically much less high- than low-fidelity data.
In order to extract as much information as possible from all available data, we here introduce an approach which aims to improve the quality of the data through denoising. We investigate the possibilities that it offers in the case of the prediction of the band gap relying on both limited experimental data and density-functional theory relying different exchange-correlation functionals (with an increasing amount of data as the accuracy of the functional decreases).
We explore different ways to combine the data into training sequences and analyze the effect of the chosen denoiser.
Finally, we analyze the effect of applying the denoising procedure several times until convergence.
Our approach provides an improvement over existing methods to exploit multi-fidelity data.

\end{abstract}


\section{Introduction}

With the considerable increase in available data, chemistry and materials science are undergoing a revolution as attested by the multiplication of data-driven studies in recent years\cite{himanen2019data, lusher2014data, Gomez-Bombarelli2018auto, schmidtRecentAdvancesApplications2019, choudharyRecentAdvancesApplications2022}.
Among all these investigations, the prediction of properties from the atomic structure (or sometimes just the chemical composition) is an extremely important topic\cite{dunn2020benchmarking}.
Indeed, having an accurate predictor can be very useful for accelerating high-throughput material screening\cite{cao2020artificial, pyzer2015what}, generating new molecules\cite{Gomez-Bombarelli2018auto}, or classifying reactions\cite{ghiandoni2019development}.
To develop such predictors, different kinds of structural descriptors have been used relying on the Coulomb matrix\cite{Rupp2012}, graphs\cite{Tsukabi2018}, voxels\cite{Kuzminykh20183d}, and so on.
Furthermore, various advanced machine learning techniques have been employed in the predictor model, such as \textit{attention}\cite{wang2021compositionally_crabnet}, \textit{graph convolution}\cite{Tsukabi2018}, \textit{embedding}\cite{Chen2019} or \textit{dimensionality reduction}\cite{debreuck2021materials}.
However, the characteristics of the data itself are seldom discussed.

In particular, many databases consist of simulated data, hence intrinsically containing some noise with respect to the experimental results (considered as ground truth here). Indeed, it is basically always necessary to resort to approximations to limit the computational power required to solve the quantum mechanical equations describing a molecule or a solid.
For instance, density-function theory (DFT) relies on an approximate functional to model the exchange-correlation energy (see e.g., Ref.~\citenum{Maurer2019}).
This necessarily comes at the cost of a reduction of the accuracy of the predicted properties.
For instance, it is well known that properties calculated within DFT may present systematic errors\cite{perdew1983physical, hautier2012accuracy_bartel15, bartel2019therole_bartel16}.
Nonetheless, many computational works focus on the relative property values for different structures, so that the systematic errors can just be canceled and the trend between the structures is not affected\cite{Bartel2020}.
In contrast, if the focus is on the absolute property value for a given structure, the difference between the experimental and calculated values cannot be neglected.
For instance, for the band gap of solids, the DFT calculations typically lead to a systematic underestimation of 30 \% - 100 \% with respect to the experimental results\cite{morales2017}.
If the experimental data is considered as the true value (even though it may vary depending on the technique), such systematic errors in the computed data meet the definition of \textit{noise}~\cite{geman1992neural}.
Note that, in the same article, the interested reader may find the definitions of \textit{bias} and \textit{variance}, as well as the relationship between those three quantities called 'bias-variance decomposition'.

In fact, the available data for one property often presents various levels of accuracy due to the different approximations adopted.
All kinds of properties have actually been studied adopting a multi-fidelity approach, such as molecular optical peaks\cite{greenman2022multi_mf1}, formation energies\cite{batra2019multifidelity_mf2, egorova2020multifidelity_mf3}, and band gap\cite{chen2021learning} which we mainly focus in this work.
As a general rule, faster calculations lead to lower accuracy results (cost versus accuracy trade-offs).
Therefore, available databases typically contain orders of magnitude less high-accuracy results than low-accuracy ones.
Recently, \citet{chen2021learning} developed a Multi-Fidelity materials Graph Networks (MFGNet) based on MatErials Graph Network (MEGNet)\cite{Chen2019} to take full advantage of all the available data by adding an integer descriptor to indicate the fidelity state (e.g., 0 - 4 for representing PBE, GLLB-SC, HSE, SCAN and experimental data) mapped to an embedding.
In the same line of thought, various methods have also been considered, including information fusion algorithms\cite{batra2019multifidelity_mf2}, Bayesian optimization\cite{egorova2020multifidelity_mf3, tran2020multi_mf4} and directed message passing neural networks\cite{greenman2022multi_mf1}. 

In the present work, we take a different route and focus on decreasing the noise on the data (i.e., reducing the input errors of the different models), taking inspiration from O2U-Net, a recent effort for improving the model performance in image classification\cite{Huang_2019_ICCV_o2u}.
The concept of noisy data dates back to the 80s and reasonable strategies were developed to handle labeling errors, provided that these affect a limited amount of the samples\cite{erkki1980_first_noise,angluin1988}.
A straightforward approach is to first remove as much noise as possible, and then train the model with the \textit{cleaned} dataset\cite{han2018coteaching_o2u5}. 
Some other works rely on curriculum learning\cite{Bengio2009} to gradually train the model using the complete dataset ordered in a meaningful sequence\cite{sheng2018curriculumnet_o2u4, jiang2017mentornet_o2u7}.

Theoretically, an additive normal distribution noise ($N\sim(\mu, \sigma^2)$) can be denoised effectively by soft-thresholding with Stein's Unbiased Risk Estimate (SURE)\cite{donoho1995noising, donoho1994ideal,sure_example}. However, the quantum chemistry systematic error is probably more like a combination of additive noise and multiplicative noise (see below).
It is thus interesting to design a new way to decrease the noise.
As the best of our knowledge, this is a first attempt in chemistry to improve the machine learning model performance in this way.

\section{Method}

\subsection{Data analysis}

In this paper, we use the same band gap datasets (four with DFT predictions and one with experimental measurements) as in Ref.~\citenum{chen2021learning}.
The four DFT datasets consist of calculations performed with the Perdew-Burke-Ernzerhof (PBE)~\cite{perdew1996generalized_pbe}, Heyd-Scuseria-Ernzerhof (HSE)~\cite{heyd2003hybrid_hse,jie2019_hse1}, strongly constrained and appropriately normed (SCAN)~\cite{sun2015strongly_scan}, and Gritsenko-Leeuwen-Lenthe-Baerends (GLLB)~\cite{gritsenko1995self_gllbsc,kuisma2010kohnsham_gllbsc2} exchange-correlation functionals for 
52348~\cite{jain2013commentary_mp}, 6030~\cite{jie2019_hse1}, 472~\cite{borlido2019large_scan1}, and 2290~\cite{castelli2015new_gllbsc3} crystalline compounds from the Materials Project, respectively.
For the sake of simplicity, these datasets will be referred to as P, H, S, and G, respectively (i.e., using the first letter of the corresponding functional).
The experimental dataset (referred to as E) comprises the band gaps of 2703 ordered crystals~\cite{zhuo2018predicting}, out of which 2401 could be assigned a most likely structure from the Materials Project~\cite{kingsbury2022}.

\begin{figure}[!h]
\includegraphics[width=0.75\textwidth]{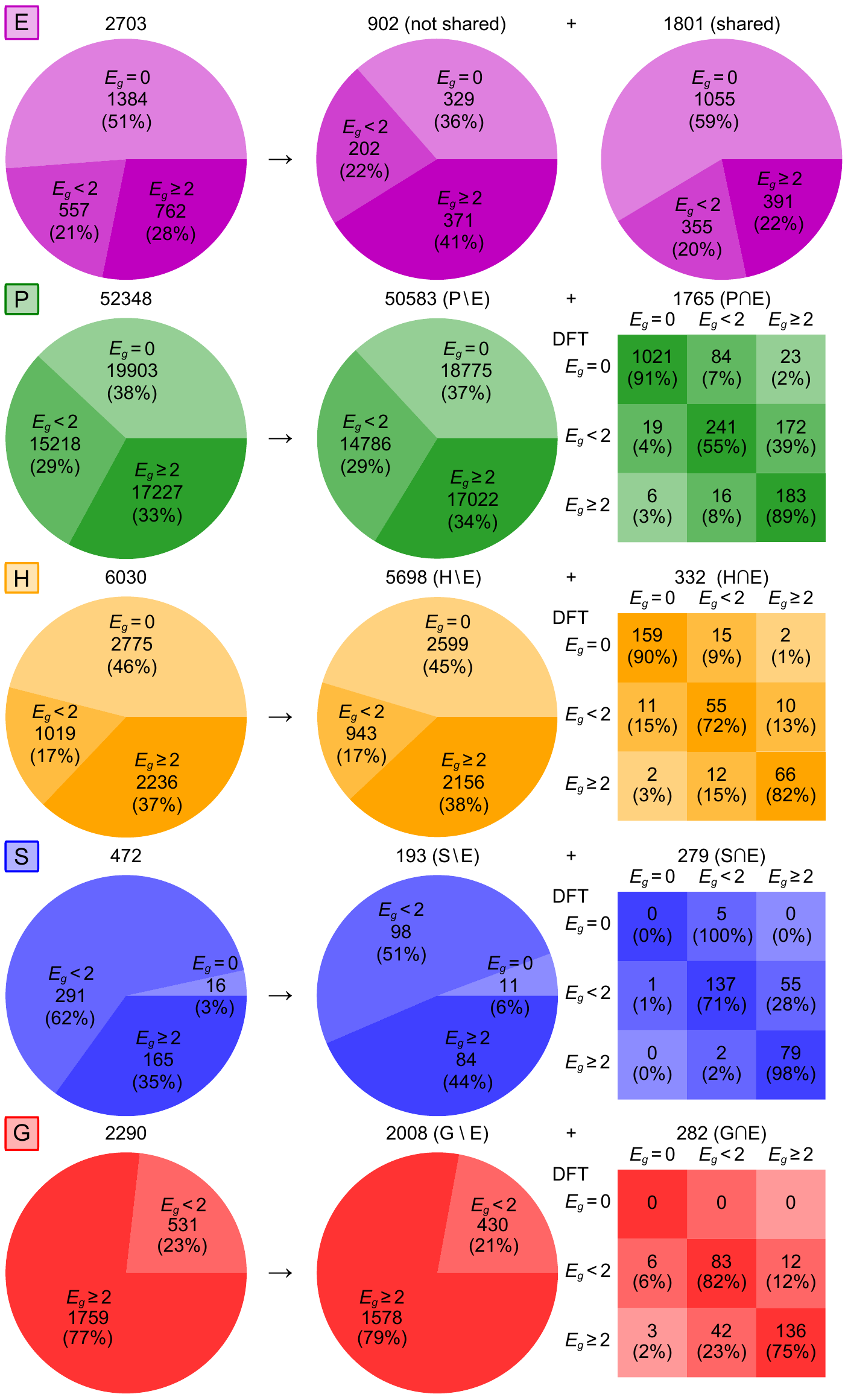}
\caption{Distribution of the data and confusion matrix as a function of the gap ($E_g$) for the different sets (E in purple, P in green, H in orange, S in blue, and G in red).}
\label{fig:piechart}
\end{figure}

As a starting analysis, it is interesting to compare how DFT (P, H, S, and G) performs with respect to the experiments (E), considered as the true values.
This can obviously only be done for the compounds that belong to the intersections between each pair of datasets (P$\cap$E, H$\cap$E, S$\cap$E, and G$\cap$E).
The resulting MAEs are reported in Table~\ref{tab:mae_dft}.
For a deeper understanding, we also indicate the MAEs corresponding to three categories of compounds: metals, as well as small-gap ($E_g$<2) and wide-gap ($E_g\geq$2) semiconductors.
In Fig.~\ref{fig:piechart}, we provide complementary information about the different datasets based on this decomposition.
For the experimental dataset (E), three piecharts have been produced with the counts and fractions of the compounds belonging to the three categories.
The first piechart concerns the whole dataset, the second one is dedicated to the data that is not shared with the DFT datasets (P, H, S, and G), and the third one focuses on the data shared with at least one of these.
For each of the latter datasets, two similar piecharts have been generated for all the data and for the part that is not shared with the dataset E while a confusion matrix has been produced for the intersections mentioned above.

\begin{table}[!htbp]
\caption{Mean absolute error (MAE) of the DFT predictions based on the different XC functionals compared with experiments.
The results are also separated between metals ($E_g$=0), small-gap ($E_g$<2) and wide-gap ($E_g\geq$2) semiconductors.
The number of compounds in each subset (\#) as well as the corresponding fraction (\%) are also reported.}
\label{tab:mae_dft}
\begin{tabular}{ccrlrrrlrrrlrrr}
\hline
  &
  \multicolumn{2}{c}{Global} & &
  \multicolumn{3}{c}{$E_g$=0} & &
  \multicolumn{3}{c}{$E_g$<2} & &
  \multicolumn{3}{c}{$E_g$$\geq$2} \\
  & MAE & \# & & MAE & \# & \% & & MAE & \# & \% & & MAE & \# & \%\\
\cline{1-3} \cline{5-7} \cline{9-11} \cline{13-15}
P$\cap$E & 0.43 & 1765 & & 0.03 & 1046 & 59.3 & & 0.61 & 341 & 19.3 & & 1.37 & 378 & 21.4 \\ 
H$\cap$E & 0.43 &  332 & & 0.07 &  172 & 51.8 & & 0.60 &  82 & 24.7 & & 1.05 &  78 & 23.5 \\ 
S$\cap$E & 0.77 &  279 & & 1.73 &    1 &  0.4 & & 0.48 & 144 & 51.6 & & 1.08 & 134 & 48.0 \\ 
G$\cap$E & 0.90 &  282 & & 2.95 &    9 &  3.2 & & 0.70 & 125 & 44.3 & & 0.94 & 148 & 52.5 \\
\hline
\end{tabular}
\end{table}

It is clear that the dataset E contains an important fraction of metals (51\%).
The accuracy of the global predictions (as measured by the MAE) is thus very sensitive to the accuracy for metals.
It turns out that PBE and HSE are doing a very good job for metals with 91\% and 90\% accuracy, respectively (see confusion matrix in Fig.~\ref{fig:piechart}, leading to a MAE of 0.03 and 0.07~eV, respectively.
In contrast, SCAN and GLLB are doing a rather poor job for metals.
For the small-gap ($E_g$<2) semiconductors, all the functionals have a very similar accuracy and MAE.
For the wide-gap ($E_g\geq$2) semiconductors, PBE clearly provides the worst prediction while the other three functionals have roughly the same predicting power.

Another important remark is that the distribution of compounds between the three different categories varies for the different intersections (P$\cap$E, H$\cap$E, S$\cap$E, and G$\cap$E).
In P$\cap$E and H$\cap$E, it is not too different from the actual distribution in the dataset E.
That is clearly not the case for S$\cap$E and G$\cap$E in which metals are strongly underrepresented.
Furthermore, in G$\cap$E, the wide-gap ($E_g\geq$2) semiconductors are largely overrepresented.
In fact, this dataset was created to analyze how the GLLB functionals performs for correcting the systematic underestimation of the band gap.

This remark is also important in the framework of the machine learning training process.
Indeed, a basic underlying assumption of such approaches is that the training dataset has a similar distribution to the test dataset.
This is a reasonable assumption for the dataset H and to a lesser extent for the dataset P, but not at all for the datasets S and G.
Given that the whole point here (and of multi-fidelity approaches) is to take advantage of all available data to overcome the lack of experimental data, we have to accept to deal with datasets with all kinds of distributions.
But it is clear that the underlying distribution will impact the ML models.

\begin{figure}[H]
\includegraphics{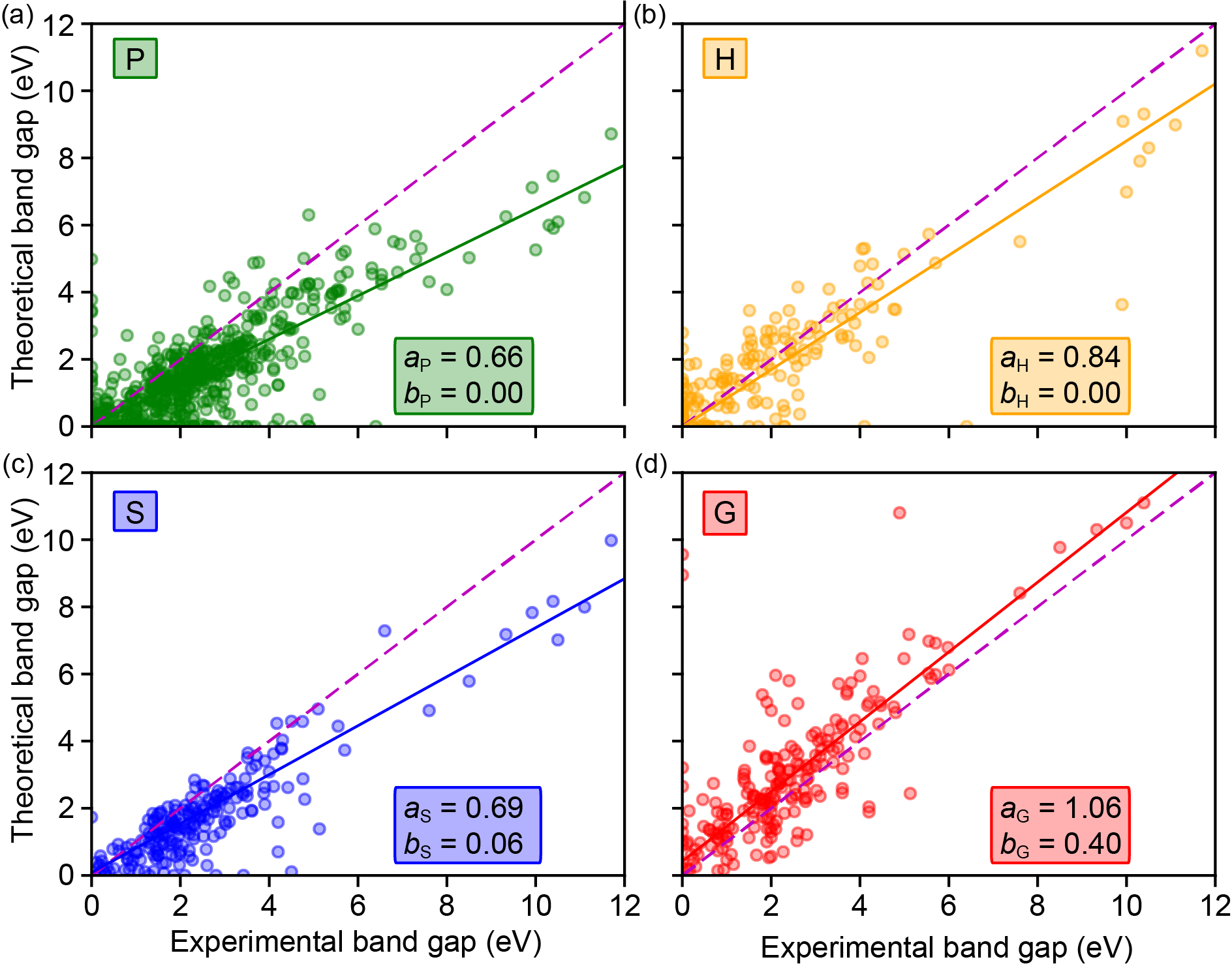}
\caption{Approximate relationship between the experimental and DFT band gaps for the different XC functionals. A first-order polynomial $P = aT + b$ is used to fit the true $T$ and predicted $P$ data points. A perfect match should lead to a slope $a = 1$ and an intercept $b = 0$. The slopes (multiplicative noise) are ordered as follows: $a_\mathrm{P} < a_\mathrm{S} < a_\mathrm{H} < 1 < a_\mathrm{G}$. The intercept (additive noise) is much bigger larger for G than for all the other XC functionals.
The color scheme is the same as in Fig.~\ref{fig:piechart}}
\label{fig:dft_exp_fitting}
\end{figure}

As far as noise is concerned, it has been long known that DFT predictions present systematic deviations.
Methods for analyzing these errors have therefore already been considered previously~\cite{lejaeghere2014error}.
The simplest model for such errors assumes that a perfect correlation exists between experimental true value $T$ and the theoretical prediction $P$, possibly with a random distortion $\epsilon$ centered around a zero mean: $P = a T + b + \epsilon$.
The parameters $a$ and $b$ account for the multiplicative and additive errors, respectively.
They can be determined by a linear regression (here, it is performed in such a way as to minimize the MAE, just as it will be done for the ML models, and a 10-fold cross-validation is used).
Fig.~\ref{fig:dft_exp_fitting} shows the results of such an analysis for the different intersections (P$\cap$E, H$\cap$E, S$\cap$E, and G$\cap$E).
The slopes for different DFT datasets are ordered as follows:
$a_\mathrm{P} < a_\mathrm{S} < a_\mathrm{H} < 1 < a_\mathrm{G}$.
Inverting the above relation between $P$ and $T$ has been suggested as a practical way to obtain improved band gaps predictions from DFT results~\cite{morales2017}: $T = c P + d + \epsilon$.
The resulting MAEs after such corrections are reported in Table~\ref{tab:mae_dft_corr}.
They are improved compared to those of the original DFT results from Table~\ref{tab:mae_dft}, mainly due to the better prediction for the wide-gap compounds.
In contrast, due to the imbalance in the distribution highlighted above, the predictions for the metals are actually worse for the dataset S (whose intersection with the dataset E only contains 1 compound which has a limited impact on the linear fit).

\begin{table}[!htbp]
\caption{Mean absolute error (MAE) of the corrected DFT predictions based on the different XC functionals compared with experiments.
The results are also separated between metals ($E_g$=0), small-gap ($E_g$<2) and wide-gap ($E_g\geq$2) semiconductors.}
\label{tab:mae_dft_corr}
\begin{tabular}{cclclclc}
\hline
  &
  Global & & $E_g$=0 & & $E_g$<2 & & $E_g$$\geq$2 \\
\cline{1-2} \cline{4-4} \cline{6-6} \cline{8-8}
P$\cap$E & 0.36 & & 0.04 & & 0.64 & & 0.99 \\ 
H$\cap$E & 0.43 & & 0.07 & & 0.60 & & 1.06 \\ 
S$\cap$E & 0.58 & & 2.31 & & 0.42 & & 0.75 \\ 
G$\cap$E & 0.67 & & 2.36 & & 0.48 & & 0.72 \\
\hline
\end{tabular}
\end{table}

In the Supporting Information, the interested reader will find further analysis of the data including the elements distribution (Sec.~\ref{sec:elements_analysis}), the overlaps between the different datasets (upset plot and Venn diagram are available in Sec.~\ref{sec:venn_upset_analysis}), the overlap plotting (in Sec.~\ref{sec:pca}) and KL divergence (Sec.~\ref{sec:KL}) of dimensionality reduction. And finally the distributions of the predicted band gap compared to the experimental values.
In other words, we examine the accuracy of the different functionals (discussed in Sec.~\ref{sec:dft_distr}).

\subsection{Training approaches and testing procedure}

It is important to remind that the experimental values are assumed to be the \textit{true values}.
The optimal ML model should thus produce results as close as possible to the experimental values, despite these may also contain some errors.
When training models, the main task is to minimize the mean absolute error (MAE) between its outputs and the true values.
Here, we adopt a twofold training-testing procedure.
The dataset E is randomly divided into two parts, E1 and E2.
The former is first used for training and the latter for testing, and then vice versa.
The final results are obtained as the average value of these two tests.
All the cases when MEGNet produces a Not-a-Number (NaN) error in one of the training folds are not considered in the analysis of the results.

\begin{figure}[H]
\includegraphics{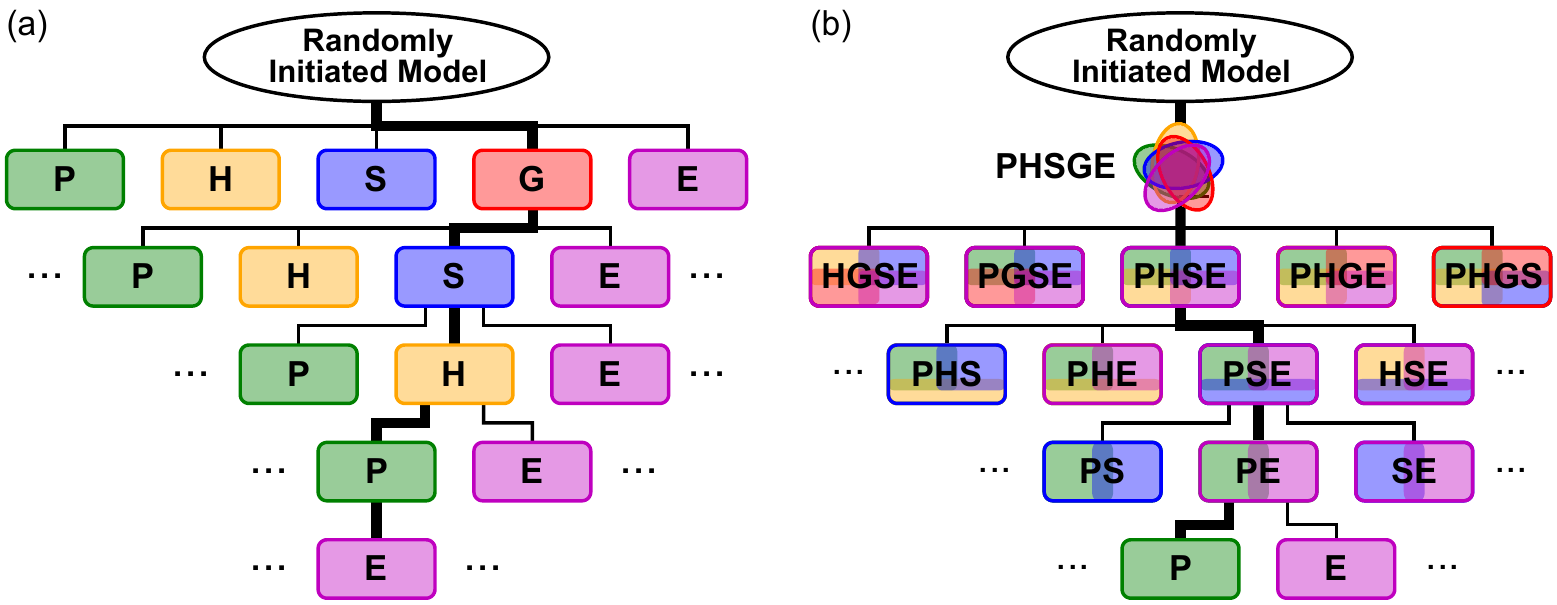}
\caption{
Schematic representation of part of the tree for (a) the \textit{one-by-one} and (b) \textit{onion} approaches. Both involve five training steps.
The thicker line in each diagram shows one possible training sequence.
For five datasets, each tree contains 120 (=5!) different branches and 325 (resp. 206) nodes in the \textit{one-by-one} (resp. \textit{onion}) approach.
Each node in the tree represents a training step.
As the datasets P, H, S, G, and E1/E2 appear (resp. disappear) with the same probability in the trees, there are 325/5=61 (resp. 206-205/5=165) nodes containing P (or any other letter) in the \textit{one-by-one} (resp. \textit{onion}) tree.}
\label{fig:tree_train_schema}
\end{figure}

In this paper, we compare four different training approaches:
\begin{enumerate}
    \item \textit{only-E}: each model is trained only on the experimental data (the first on E1 and the second on E2);
    \item \textit{all-together}: each model is trained on the combination of all five datasets (P, H, S, G, and E1/E2) regardless of their different accuracy;
    \item \textit{one-by-one}: each model is trained successively on the five datasets (P, H, S, G, and E1/E2) one at a time (so that five subsequent training steps are needed) in a selected sequence (e.g., G  $\rightarrow$ S  $\rightarrow$ H  $\rightarrow$ P  $\rightarrow$ E as illustrated in Fig.~\ref{fig:tree_train_schema}(a)).
    \item \textit{onion}: each model is trained successively on five different datasets consisting of, first, the combination of all five datasets (P, H, S, G, and E1/E2) and, then, those obtained by removing one dataset at a time in a selected sequence (e.g., PHSGE $\rightarrow$ PHSE $\rightarrow$ PSE $\rightarrow$ GE $\rightarrow$ E as illustrated in Fig.~\ref{fig:tree_train_schema}(b)).
\end{enumerate}
The \textit{only-E} approach is basically a twofold cross-validation approach that could have been used if only experimental data were available.
The \textit{one-by-one} approach is a form of curriculum learning.

There are 120 (=5!) different possible sequences for both the \textit{one-by-one} and \textit{onion} training approaches.
In this work, we consider all those alternatives systematically.
These can be represented as a tree, a part of which is shown in Fig.~\ref{fig:tree_train_schema}, highlighting one potential choice.
In what follows, we adopt the Environment for Tree Exploration (ETE) Toolkit~\cite{huerta2016ete} to display the complete tree of the different results.
By investigating all those options, which is very time consuming, we aim to analyze the sensitivity of the methods to the selected sequence.
Ideally, one would like to avoid to take them all into account for actual ML problems.
It is thus important to devise a method that is as little sensitive as possible to the selected sequence.

Note that the \textit{all-together} approach is the first step of the \textit{onion} tree, while the \textit{only-E} approach is the first step of a part of the \textit{one-by-one} tree.

\subsection{Denoising procedure}
\label{sec:denoising}

\begin{figure}[H]
\includegraphics[width=0.95\textwidth]{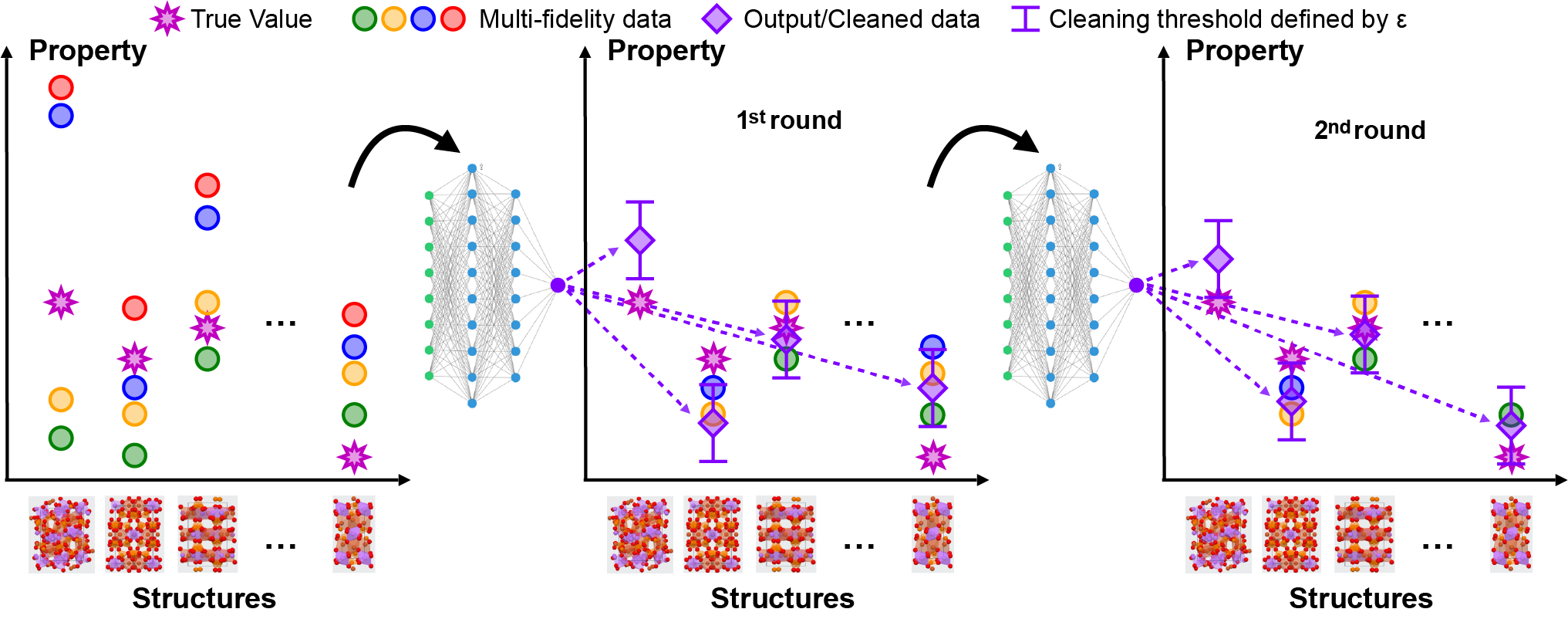}
\caption{Schematic representation of a typical data cleaning process. Target values which are too far (i.e., outside the interval defined by the cleaning threshold $\epsilon$) from the output of the predictive model are replaced by the latter values. The data cleaning is an iterative process, so more cleaning steps can be optionally added.
}
\label{fig:data_cleaning_schema}
\end{figure}

If $T$ is the target value (which includes noise) for a given sample and $P$ is the corresponding prediction by a reasonable model.
A denoising procedure typically consists in replacing $T$ by $\hat{T}=f(T,\ P)$ where $f$ is any function of $T$ and $P$ and is usually referred to as the \textit{denoiser}. Note that this can be an iterative procedure.
Given that the type of noise in our DFT datasets (typically a combination of additive noise and multiplicative noise), it is not obvious to select an existing denoiser.
In this work, we adopt a rather straightforward one:
\begin{equation}
\hat{T}=
\left\{
\begin{array}{l}
 T\textrm{\ \ if } \left|T-P\right| \leq\ \epsilon \\
 P\textrm{\ \ otherwise}
\end{array}
\right.
\label{eq:denoiser_func}
\end{equation}
where $\epsilon$ is a hyper-parameter to be determined (e.g., by grid search, random search, or Bayesian optimization).
Here, we found $\epsilon$=0.3 to be a good choice.
The whole denoising process is schematically represented in Fig.~\ref{fig:data_cleaning_schema}.

We are still left with the choice of the reasonable model to be used for making the prediction $P$.
Since the sequence of the training datasets in the \textit{one-by-one} and \textit{onion} approaches affects the final performance of the model, we tested some representative models among all the possibilities to provide an approximate error bar accounting for the denoising effect.
It is important to note that the reasonable model can be updated in an iterative process which improves the model performance until convergence is achieved.

\subsection{Machine Learning model}

For the sake of comparison with the multi-fidelity approach of \citet{chen2021learning}, we first adopt MEGNet~\cite{Chen2019} to model the relation between the structure and the band gap.
Single-fidelity models are developed for all the different datasets, using the default hyper-parameters of MEGNet (version 1.2.3).
In a second step, we also use MODNet~\cite{debreuck2021materials} (version 0.1.12) to validate our observations about the effect of denoising.

\section{Results and discussion}

\subsection{Training on the raw data}
\label{sec:training_on_raw_data}

\begin{figure}[!htbp]
\includegraphics[width=0.9\textwidth]{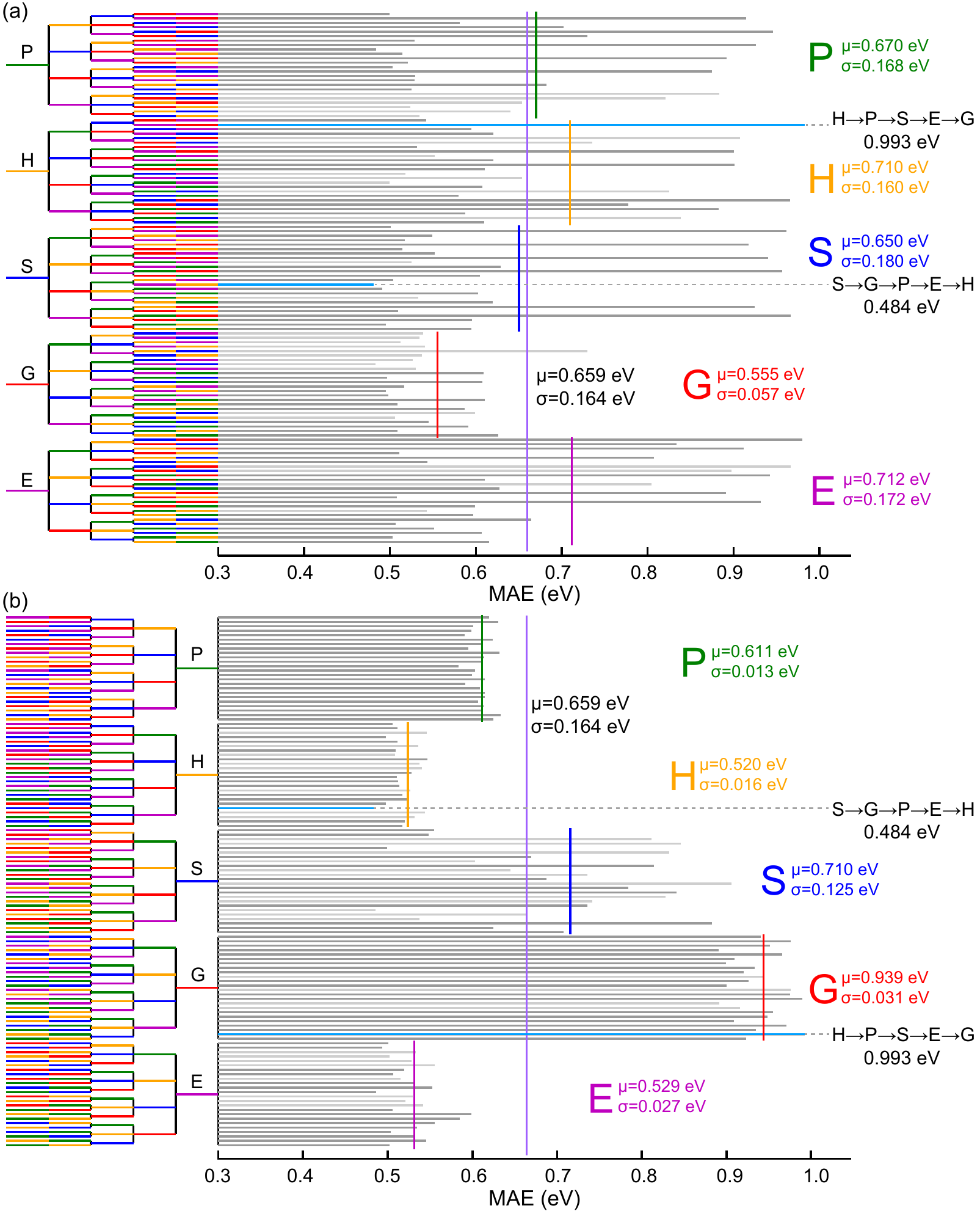}
\caption{MAE results obtained on the raw data using the \textit{one-by-one} training approach for all possible dataset sequences: (a) gathered according to the first dataset used and (b) grouped following the last dataset used.
The global average of the MAE is shown by a vertical solid purple line ($\mu$=0.659~eV), while the group averages are indicated by their corresponding color (P in green, H in orange, S in blue, G in red, and E in magenta).
The corresponding standard deviations ($\sigma$) are also indicated accordingly.
The best and worst training sequences are highlighted in light blue.
The training sequences that produce NaN for one of the folds (so the MAE is only that of the other fold) are indicated by a lighter gray bar, while those that lead to NaN for both folds are left blank.}
\label{fig:raw_data_1by1_tree_training}
\end{figure}

\begin{figure}[!htbp]
\includegraphics[width=1.0\textwidth]{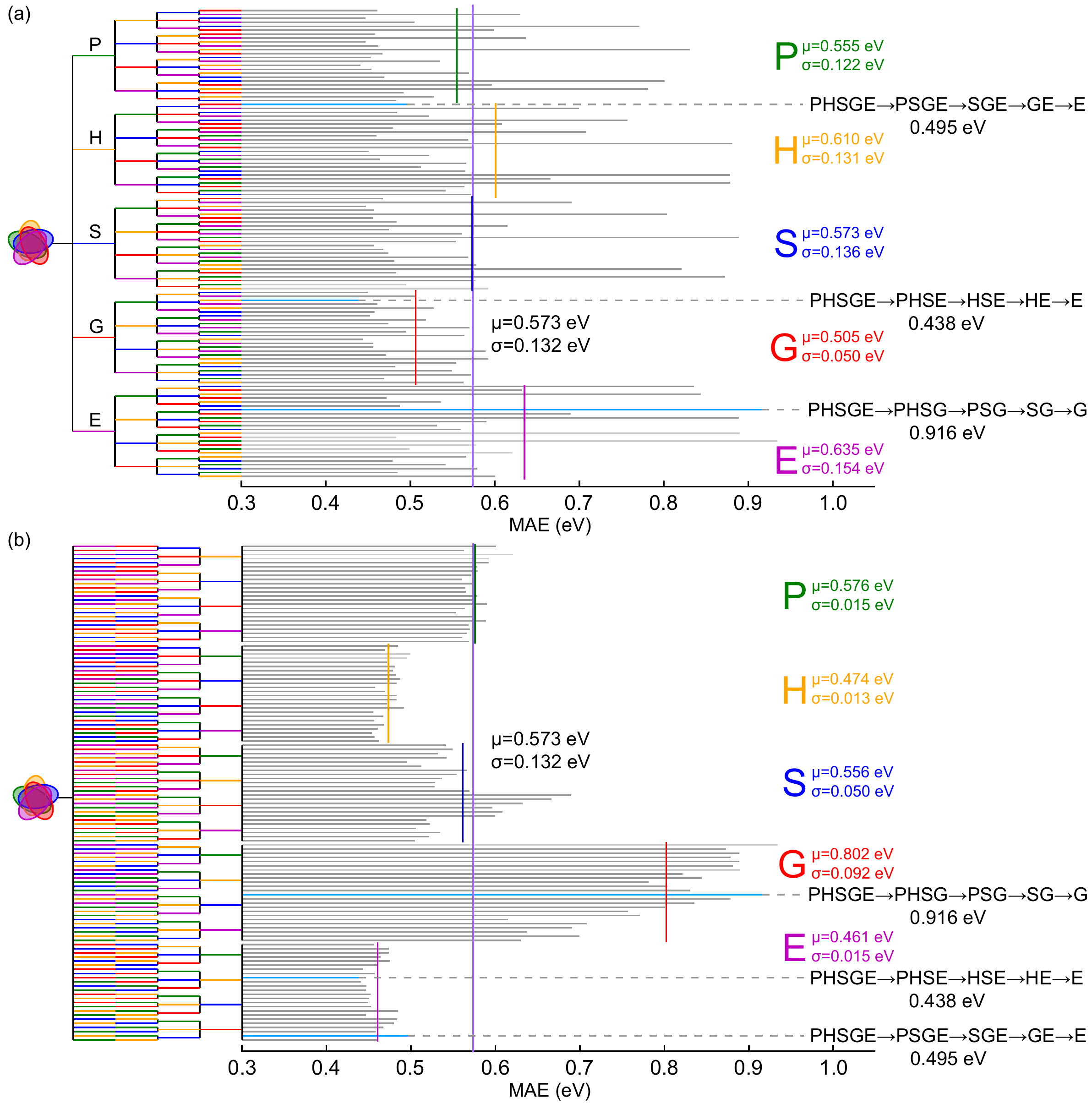}
\caption{MAE results obtained on the raw data using the \textit{onion} training approach for all possible dataset orders: (a) gathered according to the first dataset used and (b) grouped following the last dataset used.
The global average of the MAE is shown by a vertical solid purple line ($\mu$=0.573~eV), while the group averages are indicated by their corresponding color (P in green, H in orange, S in blue, G in red, and E in magenta).
The corresponding standard deviations ($\sigma$) are also indicated accordingly.
The best and worst training sequences, as well as the worst one ending by E, are highlighted in light blue.
The training sequences that produce NaN for one of the folds (so the MAE is only that of the other fold) are indicated by a lighter gray bar, while those that lead to NaN for both folds are left blank.}
\label{fig:raw_data_onion_tree_training}
\end{figure}

We first test the different approaches on the raw data (i.e., without applying the denoising procedure).
On a NVIDIA Tesla P100 graphics card, one \textit{onion} tree training costs about 6 days while the \textit{one-by-one} tree training costs about 35 days.
The most representative results are summarized in Table~\ref{tab:no_data_clean_results}, while the complete results of the \textit{one-by-one} and \textit{onion} approaches are shown in Figs.~\ref{fig:raw_data_1by1_tree_training} and \ref{fig:raw_data_onion_tree_training}.

The \textit{only-E} approach is the reference scenario.
It leads to a MAE of 0.680~eV which is higher than most of the results obtained with any other approach.
This can be traced back to the small dataset size.

The \textit{all-together} approach leads to a MAE of 0.501~eV.
That is a significant improvement by 26\%, which can be attributed to a better prediction of metals thanks to the much larger size of the dataset.
This can be understood by analyzing the results obtained by training only on the dataset P (\textit{only-P}).
This approach leads to a MAE of 0.595~eV, which is already an improvement by 13\% compared to the \textit{only-E} approach despite the fact that PBE is known to underestimate the band gap.
In fact, 72\% of the experimental data points correspond to a band gap lower than 2~eV and 51\% are actually metals.
If we focus on the intersection P$\cap$E (containing 1765 compounds), we see that 59\% of the compounds are metallic and P is actually correct in 91\% of the cases.
The underestimation of the band gap only leads to 9\% of false metallic compounds.
Now, moving to the rest of the datasets P (P$\setminus$E), we see that, out of the 50583 compounds, 18775 (37\%) are metals.
This number is basically one order of magnitude larger than the 1384 metallic compounds present in the whole dataset E.
So, the ML model can better learn to predict metals.
Adding the fact that another 15218 compounds have a band gap smaller than 2~eV for which the PBE error is not going to be very big, we can easily understand the nice improvement in MAE.
For the datasets S and G, the number of new metallic systems added compared to E (11, and 0, respectively) is much smaller.
So, not surprisingly, \textit{only-S} and \textit{only-G} suffer much more from the noise due to the XC functionals than the \textit{only-P} one leading a MAE of 1.446 and 1.406~eV, respectively.
For the \textit{all-together} approach, the improvement results from both the effect of the number of metallic samples and an averaging of the noise of the different XC functional.
The \textit{only-H} results are somewhere in between with a MAE of 0.796~eV.
Indeed, the number of new metals in H$\setminus$E (2599) is only the double than in E (compared to more than 10 times in P).
So, the effect of the better prediction of metals is more limited compared to P.

For the \textit{one-by-one} and \textit{onion} approaches, the results vary depending on the training sequence.
The best and worst results are reported in Table~\ref{tab:no_data_clean_results}.
In order to analyze the effect of the training sequence, we have produced two plots for both approaches in Figs.~\ref{fig:raw_data_1by1_tree_training} and \ref{fig:raw_data_onion_tree_training}.
In the first part of those figures, the sequences are classified according to the first dataset used or removed (P, H, S, G, or E); while, in the second part, they are ordered depending on the last dataset used.

In both figures, each class presents much more variation around its mean in the first plot than in the second one.
In other words, the final dataset used seems to matter much more than the first one used (resp. removed) in the \textit{one-by-one} (resp. \textit{onion}) approach.
It is, however, also clear that using the dataset G first leads to better results and not surprisingly finishing the training with it produces the worst results by far.
For the \textit{one-by-one} approach, the best results on average are obtained for the sequences finishing with H.
They are slightly better than those finishing with E.
For the \textit{onion} approach, it is actually the reverse: the best results on average being achieved for the sequences finishing with E.
As a general rule, in order to limit the number of models to be tested, one can clearly focus on the latter sequences (i.e., those finishing with the available true values) and, for further restriction, one can concentrate on those which end with PE or HE given that P and H have the lowest MAE (i.e., the highest fidelity) in Table~\ref{tab:mae_dft}.

\begin{table}[!htbp]
    \caption{Most representative results (in terms of the mean absolute error on the predicted bang gap expressed in eV) obtained for the different training approaches and sequences without any denoising. For the \textit{one-by-one} and \textit{onion} approaches, we report the results for the best training sequence, the worst one, and the worst one among those ending with E (indicated with a star).}
    \label{tab:no_data_clean_results}
    \centering
    \setlength{\tabcolsep}{4pt}% column separation
    \renewcommand{\arraystretch}{1.2}% row space 
    \begin{tabular}{llllll}
        \hline
                 &       & \multicolumn{4}{c}{MAE} \\
        Approach & Sequence &Global & $E_g$=0 & $E_g$<2 & $E_g$$\geq$2 \\
        \hline
        \textit{only-E} & E & 0.680 & 0.490 & 0.534 & 1.131 \\
        \textit{all-together} & PHSGE & 0.501 & 0.150 & 0.567 & 1.091 \\
        \textit{one-by-one} & S$\rightarrow$G$\rightarrow$P$\rightarrow$E$\rightarrow$H (best) & 0.484 & 0.228 & 0.571 & 0.884 \\
        & H$\rightarrow$P$\rightarrow$S$\rightarrow$E$\rightarrow$G (worst) & 0.993 & 1.039 & 0.733 & 1.097\\
        & H$\rightarrow$P$\rightarrow$G$\rightarrow$S$\rightarrow$E (worst$^*$) & 0.599 & 0.434 & 0.510 & 0.964 \\
        \textit{onion} & PHSGE$\rightarrow$PHSE$\rightarrow$HSE$\rightarrow$HE$\rightarrow$E (best) & 0.438 & 0.239 & 0.515 & 0.743 \\
        & PHSGE$\rightarrow$PHSG$\rightarrow$PSG$\rightarrow$SG$\rightarrow$G (worst) & 0.916 & 0.937 & 0.634 & 1.083 \\
        & PHSGE$\rightarrow$PHSG$\rightarrow$PSG$\rightarrow$PG$\rightarrow$G (2$^{nd}$-worst) & 0.889 & 0.883 & 0.665 & 1.064 \\
        & PHSGE$\rightarrow$PSGE$\rightarrow$SGE$\rightarrow$GE$\rightarrow$E (worst$^*$) & 0.495 & 0.338 & 0.485 & 0.790 \\
        \hline
    \end{tabular}
\end{table}

\subsection{Training with denoised data}

\begin{figure}[H]
\includegraphics[width=0.9\textwidth]{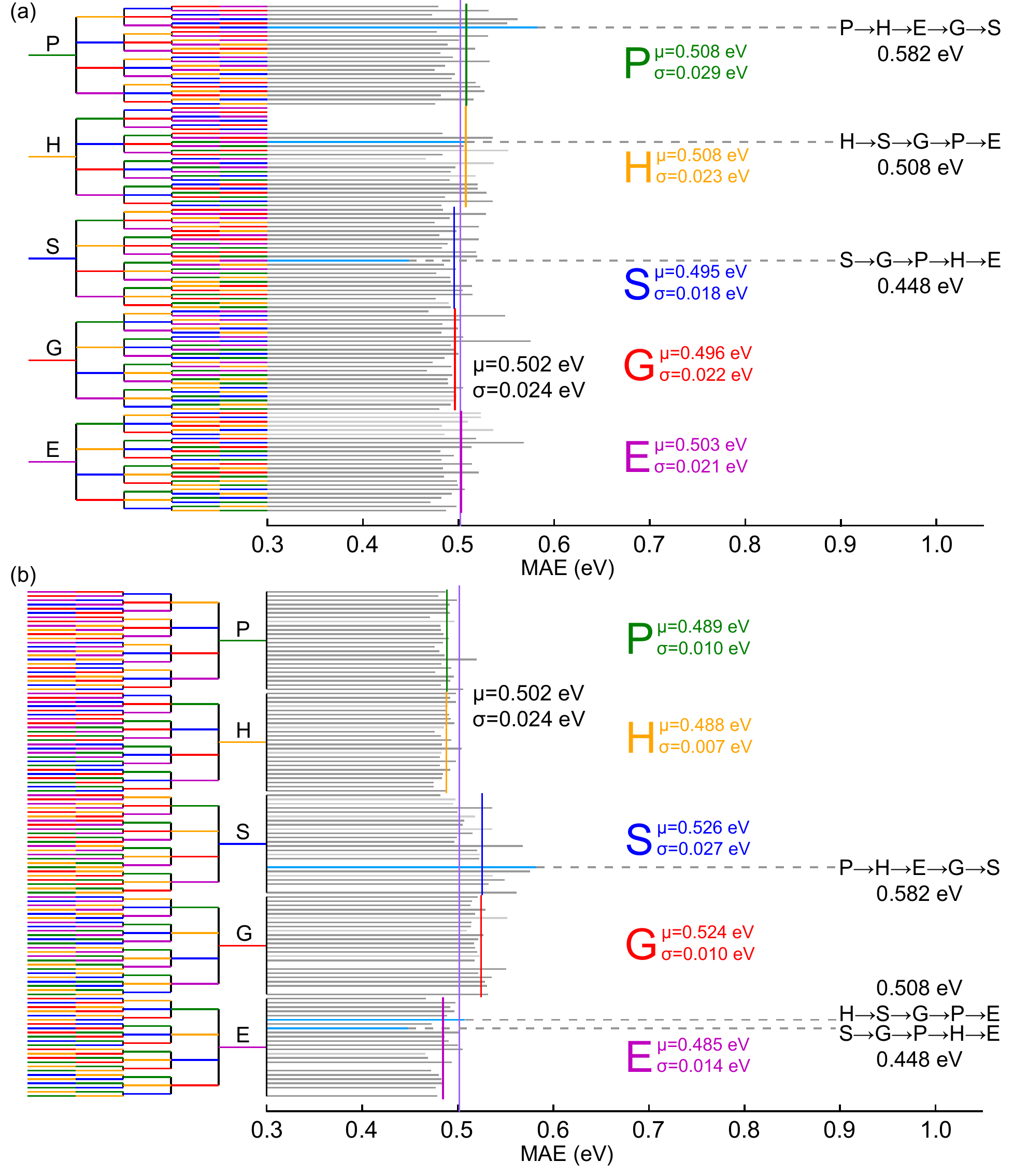}
\caption{MAE results obtained on the data cleaned with the best model of Fig.~\ref{fig:raw_data_1by1_tree_training} using the \textit{one-by-one} training approach for all possible dataset sequences: (a) gathered according to the first dataset used and (b) grouped following the last dataset used.
The global average of the MAE is shown by a vertical solid purple line ($\mu$=0.502~eV), while the group averages are indicated by their corresponding color (P in green, H in orange, S in blue, G in red, and E in magenta).
The corresponding standard deviations ($\sigma$) are also indicated accordingly.
The best and worst training sequences, as well as the worst one ending by E, are highlighted in light blue.
The training sequences that produce NaN for one of the folds (so the MAE is only that of the other fold) are indicated by a lighter gray bar, while those that lead to NaN for both folds are left blank.}
\label{fig:cleaned_data_1by1_tree_training}
\end{figure}

\begin{figure}[H]
\includegraphics[width=0.9\textwidth]{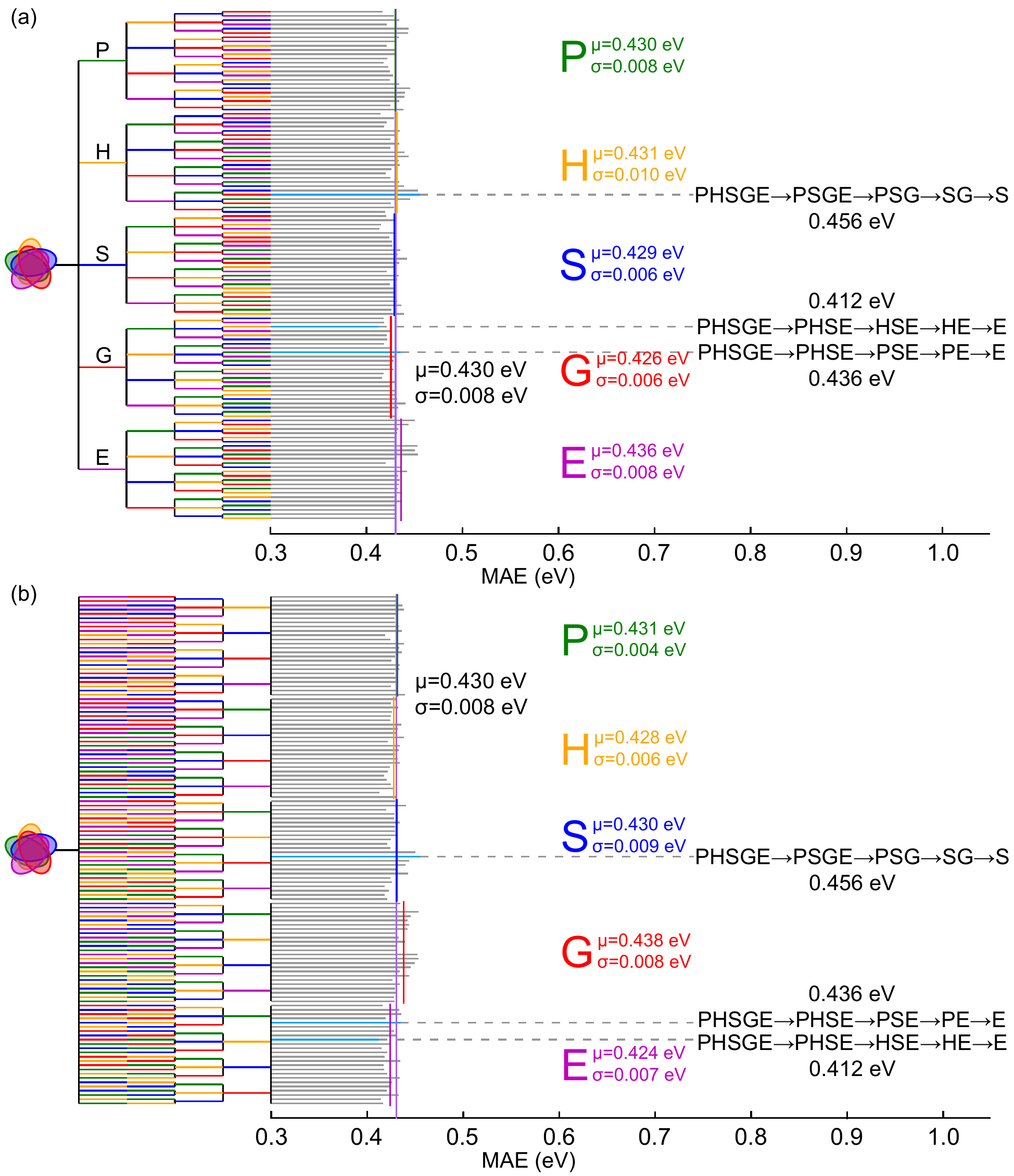}
\caption{MAE results obtained on the data cleaned with the best model (MAE=0.438~eV) of Fig.~\ref{fig:raw_data_onion_tree_training} using the \textit{onion} training approach for all possible dataset orders: (a) gathered according to the first dataset used and (b) grouped following the last dataset used.
The global average of the MAE is shown by a vertical solid purple line ($\mu$=0.430~eV), while the group averages are indicated by their corresponding color (P in green, H in orange, S in blue, G in red, and E in magenta).
The corresponding standard deviations ($\sigma$) are also indicated accordingly.
The best and worst training sequences, as well as the worst one ending by E, are highlighted in light blue.
The training sequences that produce NaN for one of the folds (so the MAE is only that of the other fold) are indicated by a lighter gray bar, while those that lead to NaN for both folds are left blank.}
\label{fig:cleaned_data_onion_tree_training}
\end{figure}

We now turn to the analysis of the results that can be obtained when denoising the data.
Given that we have already considered all the possible training sequences, the natural choice to clean the data is to use the best model obtained with the raw data.
Once again, we first analyze the effect of the training sequence.
The results obtained for both \textit{one-by-one} and \textit{onion} approaches are reported in Figs.~\ref{fig:cleaned_data_1by1_tree_training} and \ref{fig:cleaned_data_onion_tree_training}.
The striking difference with respect to the results obtained on the raw data is that the training sequence has a much smaller impact on the results.
This is a really important point in order to avoid the burden of having to compute all the different training sequences.
The second important observation is that, once again, the \textit{onion} approach produces better results than \textit{one-by-one}.
So, from now on, we focus on the \textit{onion} approach to analyze the effects of the cleaning procedure.

Given that in a normal investigation the best possible model will not be known \textit{a priori} (it only can \textit{a posteriori} once all sequences have been considered), we investigate the importance of the choice of the denoiser.
Here, we have plenty of models at hand differing by the training sequence in the raw data.
Besides the one already considered, we select four other denoiser models for comparison:
\begin{itemize}
\item
PHSGE$\rightarrow$PHSG$\rightarrow$PSG$\rightarrow$SG$\rightarrow$G which leads to the worst performance among all training paths: MAE=0.916~eV (Fig. \ref{fig:1st_worst_denoiser_data_tree_training}),
\item
PHSGE$\rightarrow$PHSG$\rightarrow$PSG$\rightarrow$PG$\rightarrow$G which leads to the second-worst performance among all training paths: MAE=0.889~eV (Fig. \ref{fig:2nd_worst_denoiser_data_tree_training}),
\item
PHSGE$\rightarrow$PHGE$\rightarrow$PGE$\rightarrow$GE$\rightarrow$E which has a rather poor performance among all training path ending with E: MAE=0.483~eV (Fig. \ref{fig:3rd_worst_of_ending_E_denoiser_data_tree_training}),
\item
PHSGE$\rightarrow$PHSE$\rightarrow$PHE$\rightarrow$HE$\rightarrow$E which has a rather good performance among all training path ending with E: MAE=0.443~eV (Fig. \ref{fig:3rd_best_EGPHS_EPHS_EPH_EH_E_denoiser_onion_tree_training}).
\end{itemize}

The complete results obtained after the denoising procedure based on these four different models are shown in Figs.~\ref{fig:1st_worst_denoiser_data_tree_training}, \ref{fig:2nd_worst_denoiser_data_tree_training},   \ref{fig:3rd_worst_of_ending_E_denoiser_data_tree_training}, and
\ref{fig:3rd_best_EGPHS_EPHS_EPH_EH_E_denoiser_onion_tree_training} in the Supporting Information (Sec.~\ref{sec:tree}).

In all four cases, the denoising procedure improves the global average of the MAE for the whole tree, as well as the average MAE of all the sequences ending with E compared to the results of the denoiser model itself.
However, when the worst or the second-worst model is used as the denoiser, the results are worse than with the raw data.

Basically, we observe that the better the denoiser model the better the cleaning effects, which translates not only into a lower MAE but also into a lower variance with respect to the training sequence.
Therefore, the choice of the denoiser is quite critical.

It would be cheating to use the final results as an indicator to choose the denoiser model. 
However, we note that, as soon as a model whose training sequence ends with E is chosen as the denoiser (even the rather poor performance one), the results are clearly improved with respect to those obtained based on the raw data.
Therefore, based on the observations at the end of Sec.~\ref{sec:training_on_raw_data}, we recommend as heuristic to use a denoiser for which the training sequence is in increasing fidelity of the data ending with the true values.

\begin{figure}[H]
\includegraphics[width=\textwidth]{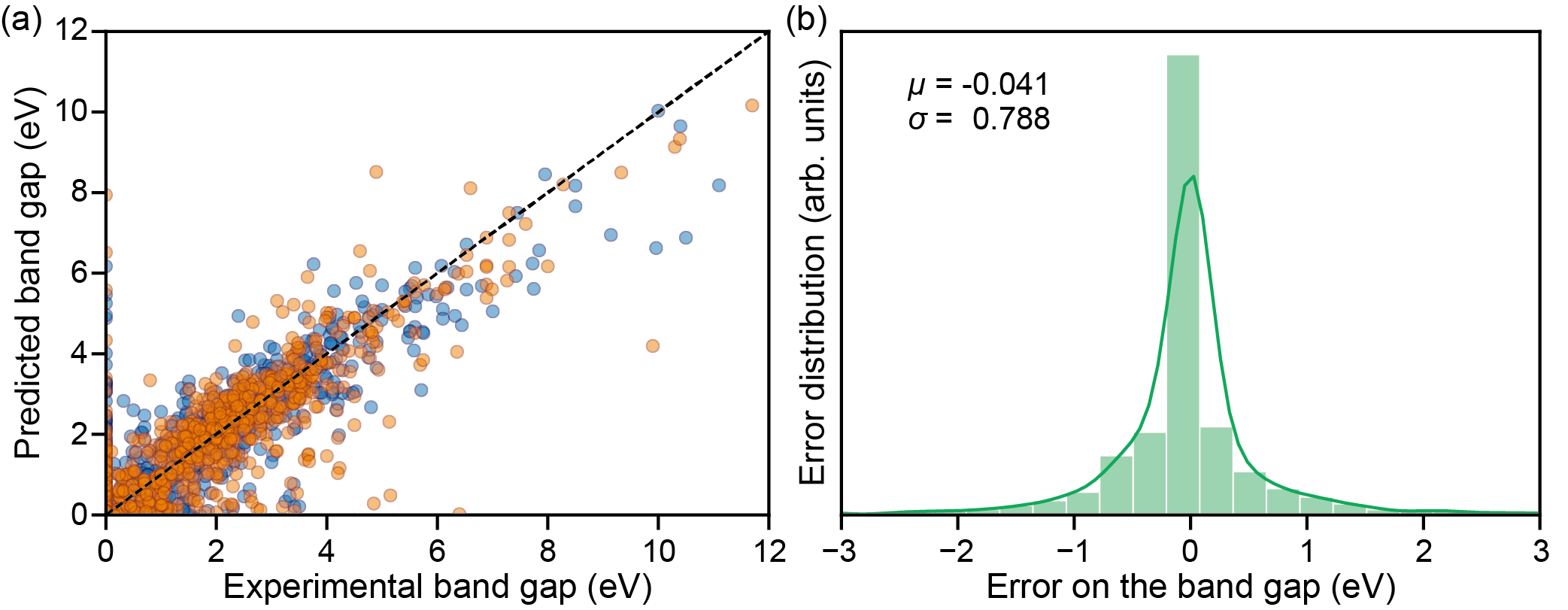}
\caption{(a) Analysis the model prediction error through scatter and slope (Fig. \ref{fig:dft_exp_fitting} style). Different colors are adopted to show the twofold results.
Analysis the model prediction error through distribution (Fig. 
(b) A typical model output vs. experimental data. All test/training are made by twofold.}
\label{fig:model_output_vs_exp}
\end{figure}

The final results obtained after the denoising procedure, together with the distribution of errors, are represented in Fig.~\ref{fig:model_output_vs_exp}.
Compared with Figs.~\ref{fig:dft_exp_fitting} and \ref{fig:band_gap_err}, it appears rather clearly that these contain less noise.

\begin{figure}[H]
\includegraphics{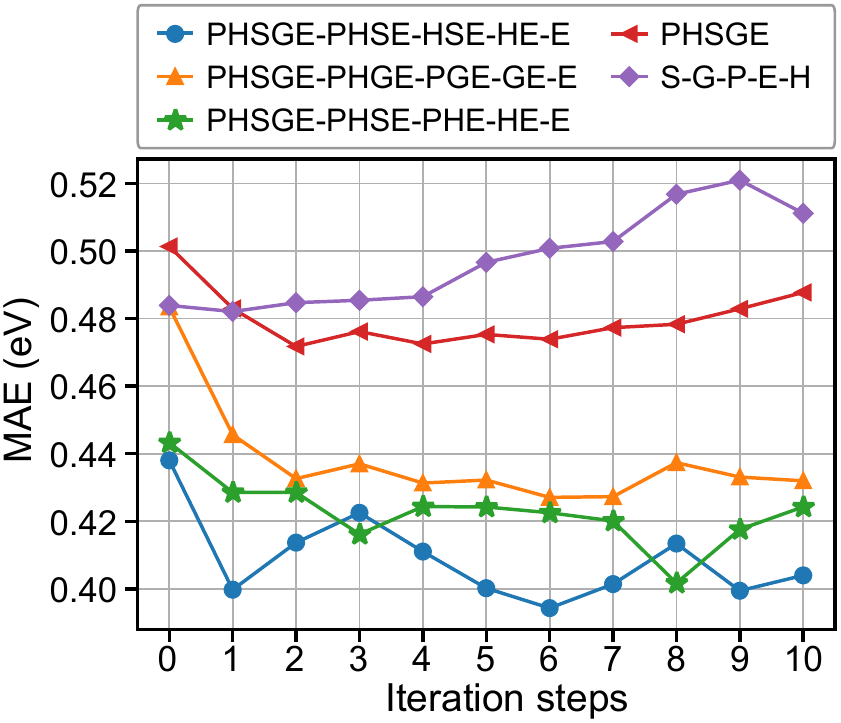}
\caption{Application of an iterative procedure for some representative training sequences.
The first point of each line is the result obtained with the raw data (i.e., without data cleaning).
Every subsequent point is obtained using the model corresponding to the previous point as the denoiser for cleaning the data.}
\label{fig:converging_effect}
\end{figure}

As already indicated, the cleaning procedure can be iterated towards convergence.
In Fig.~\ref{fig:converging_effect}, we show the evolution of the results as a function of the iteration for some representative training sequences. 
PHSGE$\rightarrow$PHSE$\rightarrow$HSE$\rightarrow$HE$\rightarrow$E) leads to the lowest MAE (0.394~eV at the 6$^{th}$ iteration). 
Compared with \textit{one-by-one} and \textit{all-together} approaches, the \textit{onion} training not only shows the best performance at the starting point, but it also has the greatest potential for improvement.
The \textit{one-by-one} training results can actually hardly be improved by the cleaning procedure due to the lack of a real synergetic effect by the different datasets.

\subsection{Validation of the denoising method with MODNet}

To assess the generality of the denoising method, we also adopt MODNet~\cite{debreuck2021materials} to model the relation between the structure and the band gap.
Indeed, it is among the best models of the MatBench test suite~\cite{dunn2020benchmarking}.

We use the same train/test splitting of the dataset as for the work already performed with MEGNet.
For the \textit{onion} approach, we adopt the training sequence that was found to be the best with MEGNet.
Here, we have no clue whether it is also the best with MODNet.
However, it meets the heuristic defined above for obtaining a reasonable denoiser.

The results are summarized in Table~\ref{tab:modnet}.
As already observed with MEGNet, the onion training and the data cleaning improve the predictions compared to training on the raw experimental data (\textit{only-E}).
Compared to MEGNet, these improvements are less impressive given that the reference results (\textit{only-E}) were already reasonably good.
As a side note, MODNet is much less computationally demanding than MEGNet (about 10 minutes vs. about 8 hours) due to the smaller model size.

\begin{table}[!htbp]
    \caption{Most representative results (in terms of the mean absolute error on the predicted bang gap expressed in eV) obtained for the different training approaches and sequences with/without denoising.}
    \label{tab:modnet}
    \centering
    \setlength{\tabcolsep}{4pt}% column separation
    \renewcommand{\arraystretch}{1.2}% row space 
    \begin{tabular}{lllcc}
        \hline
        & & &\multicolumn{2}{c}{MAE (eV)} \\
        Approach & Denoiser & Sequence & MODNet & MEGNet \\
        \hline
        \textit{only-E} & \textit{none} & E & 0.458 & 0.680 \\
        \textit{onion} & \textit{none} & PHSGE$\rightarrow$PHSE$\rightarrow$HSE$\rightarrow$HE$\rightarrow$E & 0.422 & 0.438 \\
        % \textit{onion} & \textit{only-E} & PHSGE$\rightarrow$PHSE$\rightarrow$HSE$\rightarrow$HE$\rightarrow$E & 0.427 & - \\
        \textit{onion} & \textit{onion} & PHSGE$\rightarrow$PHSE$\rightarrow$HSE$\rightarrow$HE$\rightarrow$E & 0.406 & 0.412 \\
        \hline
    \end{tabular}
\end{table}

\section{Conclusion}
In this paper, we have introduced a method to take full advantage of the availability of multi-fidelity data and tested it thoroughly for the prediction of the band gap based on the structure.
The method is based on an appropriate combination of all the data into a multistep training sequence and on a simple denoising procedure.
For combining the data, we have compared four different training approaches (\textit{only-E}, \textit{all-together}, \textit{one-by-one}, and \textit{onion}).
It turned out that the best one consists in training the model successively on different datasets resulting from, first,
the combination of all available datasets and, then, of those obtained
by removing one dataset at a time by increasing fidelity (from the poorest to the highest fidelity, hence, finishing with the true data).
For the denoising procedure, we have tested a simple technique by which target values are replaced by the output of the selected denoiser when the former are too far (i.e., outside the interval defined by a cleaning threshold) from the latter. Other denoising procedures resulting in better results might be existing, but is left for future work.
We have found that the denoising procedure improves the final results provided that a reasonable denoiser is chosen.
%Furthermore, based on our observations, we have proposed a few simple prescriptions to select an acceptable one.
Furthermore, based on our observations, we proposed a simple heuristic for the denoiser.
Finally, we have investigated the effect of applying the denoising procedure several times until convergence.

The method proposed here provides a sensible way to improve the results that can be achieved when multi-fidelity data are available which is basically often the case in materials science given that accuracy in the data always comes at a cost.
It thus has considerable potential of applications.

\section*{Acknowledgement}
X.T.L is grateful for the funding support from Beijing Advanced
Innovation Center for Materials Genome Engineering (Beijing Information Science and Technology University) and National Natural Science Foundation of China (No. 22002008).
P.-P.D.B. and G.-M.R. are grateful to the F.R.S.-FNRS for financial support.
We also thank to the authors of MEGNet and MFGNet. They inspired us and help us fix some issues on Github.
X.T.L. also thanks to Prof. Ning Li, Dr. Tao Yang, Prof. Xiaodong Wen and Mr. Enhu Diao for providing feedback and help on this work.

\section*{Author contributions}

X.T.L. and G.-M.R conceived the idea and designed the work. X.T.L. implemented the models and performed the analysis. P.-P.D.B provided the data and validated the denoising performance with MODNet. L.H.W. helped with the data analysis and figure plotting. G.-M.R supervised the project. All authors wrote the manuscript and contributed to the discussion and revision.

\section*{Competing interests}
The authors declare no competing interests.

\section*{Associated Content}

\subsection*{Supporting Information}

The source code is available on \url{https://github.com/liuxiaotong15/denoise}
for the tree training with MEGNet, on
\url{https://github.com/ppdebreuck/onion_modnet}
for MODNet validation
and on
%\url{http://chemdata.herokuapp.com/} 
\url{https://bandgap-denoiser.modl-uclouvain.org/} for comparison of all error distribution.

The Supporting Information consist of Secs.~\ref{sec:dataset_analysis}: \nameref{sec:dataset_analysis} and \ref{sec:tree}: \nameref{sec:tree}.

% Table S1-S5 and 
% Figs.~S1 \ref{fig:3rd_best_EGPHS_EPHS_EPH_EH_E_denoiser_onion_tree_training} together with previous 2 parts are available free of charge via the Internet.

\newpage
\bibliography{ref}

\providecommand{\latin}[1]{#1}
\makeatletter
\providecommand{\doi}
  {\begingroup\let\do\@makeother\dospecials
  \catcode`\{=1 \catcode`\}=2 \doi@aux}
\providecommand{\doi@aux}[1]{\endgroup\texttt{#1}}
\makeatother
\providecommand*\mcitethebibliography{\thebibliography}
\csname @ifundefined\endcsname{endmcitethebibliography}
  {\let\endmcitethebibliography\endthebibliography}{}
\begin{mcitethebibliography}{51}
\providecommand*\natexlab[1]{#1}
\providecommand*\mciteSetBstSublistMode[1]{}
\providecommand*\mciteSetBstMaxWidthForm[2]{}
\providecommand*\mciteBstWouldAddEndPuncttrue
  {\def\EndOfBibitem{\unskip.}}
\providecommand*\mciteBstWouldAddEndPunctfalse
  {\let\EndOfBibitem\relax}
\providecommand*\mciteSetBstMidEndSepPunct[3]{}
\providecommand*\mciteSetBstSublistLabelBeginEnd[3]{}
\providecommand*\EndOfBibitem{}
\mciteSetBstSublistMode{f}
\mciteSetBstMaxWidthForm{subitem}{(\alph{mcitesubitemcount})}
\mciteSetBstSublistLabelBeginEnd
  {\mcitemaxwidthsubitemform\space}
  {\relax}
  {\relax}

\bibitem[Himanen \latin{et~al.}(2019)Himanen, Geurts, Foster, and
  Rinke]{himanen2019data}
Himanen,~L.; Geurts,~A.; Foster,~A.~S.; Rinke,~P. Data-driven materials
  science: status, challenges, and perspectives. \emph{Advanced Science}
  \textbf{2019}, \emph{6}, 1900808\relax
\mciteBstWouldAddEndPuncttrue
\mciteSetBstMidEndSepPunct{\mcitedefaultmidpunct}
{\mcitedefaultendpunct}{\mcitedefaultseppunct}\relax
\EndOfBibitem
\bibitem[Lusher \latin{et~al.}(2014)Lusher, McGuire, van Schaik, Nicholson, and
  de~Vlieg]{lusher2014data}
Lusher,~S.~J.; McGuire,~R.; van Schaik,~R.~C.; Nicholson,~C.~D.; de~Vlieg,~J.
  Data-driven medicinal chemistry in the era of big data. \emph{Drug discovery
  today} \textbf{2014}, \emph{19}, 859--868\relax
\mciteBstWouldAddEndPuncttrue
\mciteSetBstMidEndSepPunct{\mcitedefaultmidpunct}
{\mcitedefaultendpunct}{\mcitedefaultseppunct}\relax
\EndOfBibitem
\bibitem[G{\'o}mez-Bombarelli \latin{et~al.}(2018)G{\'o}mez-Bombarelli, Wei,
  Duvenaud, Hern{\'a}ndez-Lobato, S{\'a}nchez-Lengeling, Sheberla,
  Aguilera-Iparraguirre, Hirzel, Adams, and
  Aspuru-Guzik]{Gomez-Bombarelli2018auto}
G{\'o}mez-Bombarelli,~R.; Wei,~J.~N.; Duvenaud,~D.;
  Hern{\'a}ndez-Lobato,~J.~M.; S{\'a}nchez-Lengeling,~B.; Sheberla,~D.;
  Aguilera-Iparraguirre,~J.; Hirzel,~T.~D.; Adams,~R.~P.; Aspuru-Guzik,~A.
  Automatic chemical design using a data-driven continuous representation of
  molecules. \emph{ACS central science} \textbf{2018}, \emph{4}, 268--276\relax
\mciteBstWouldAddEndPuncttrue
\mciteSetBstMidEndSepPunct{\mcitedefaultmidpunct}
{\mcitedefaultendpunct}{\mcitedefaultseppunct}\relax
\EndOfBibitem
\bibitem[Schmidt \latin{et~al.}()Schmidt, Marques, Botti, and
  Marques]{schmidtRecentAdvancesApplications2019}
Schmidt,~J.; Marques,~M. R.~G.; Botti,~S.; Marques,~M. A.~L. Recent Advances
  and Applications of Machine Learning in Solid-State Materials Science.
  \emph{5}, 83\relax
\mciteBstWouldAddEndPuncttrue
\mciteSetBstMidEndSepPunct{\mcitedefaultmidpunct}
{\mcitedefaultendpunct}{\mcitedefaultseppunct}\relax
\EndOfBibitem
\bibitem[Choudhary \latin{et~al.}()Choudhary, DeCost, Chen, Jain, Tavazza,
  Cohn, Park, Choudhary, Agrawal, Billinge, Holm, Ong, and
  Wolverton]{choudharyRecentAdvancesApplications2022}
Choudhary,~K.; DeCost,~B.; Chen,~C.; Jain,~A.; Tavazza,~F.; Cohn,~R.;
  Park,~C.~W.; Choudhary,~A.; Agrawal,~A.; Billinge,~S. J.~L. \latin{et~al.}
  Recent Advances and Applications of Deep Learning Methods in Materials
  Science. \emph{8}, 59\relax
\mciteBstWouldAddEndPuncttrue
\mciteSetBstMidEndSepPunct{\mcitedefaultmidpunct}
{\mcitedefaultendpunct}{\mcitedefaultseppunct}\relax
\EndOfBibitem
\bibitem[Dunn \latin{et~al.}(2020)Dunn, Wang, Ganose, Dopp, and
  Jain]{dunn2020benchmarking}
Dunn,~A.; Wang,~Q.; Ganose,~A.; Dopp,~D.; Jain,~A. Benchmarking materials
  property prediction methods: the matbench test set and automatminer reference
  algorithm. \emph{npj Computational Materials} \textbf{2020}, \emph{6},
  138\relax
\mciteBstWouldAddEndPuncttrue
\mciteSetBstMidEndSepPunct{\mcitedefaultmidpunct}
{\mcitedefaultendpunct}{\mcitedefaultseppunct}\relax
\EndOfBibitem
\bibitem[Cao \latin{et~al.}(2020)Cao, Ouyang, Ghiringhelli, Scheffler, Liu,
  Carbogno, and Zhang]{cao2020artificial}
Cao,~G.; Ouyang,~R.; Ghiringhelli,~L.~M.; Scheffler,~M.; Liu,~H.; Carbogno,~C.;
  Zhang,~Z. Artificial intelligence for high-throughput discovery of
  topological insulators: The example of alloyed tetradymites. \emph{Physical
  Review Materials} \textbf{2020}, \emph{4}, 034204\relax
\mciteBstWouldAddEndPuncttrue
\mciteSetBstMidEndSepPunct{\mcitedefaultmidpunct}
{\mcitedefaultendpunct}{\mcitedefaultseppunct}\relax
\EndOfBibitem
\bibitem[Pyzer-Knapp \latin{et~al.}(2015)Pyzer-Knapp, Suh,
  G{\'o}mez-Bombarelli, Aguilera-Iparraguirre, and Aspuru-Guzik]{pyzer2015what}
Pyzer-Knapp,~E.~O.; Suh,~C.; G{\'o}mez-Bombarelli,~R.;
  Aguilera-Iparraguirre,~J.; Aspuru-Guzik,~A. What is high-throughput virtual
  screening? A perspective from organic materials discovery. \emph{Annual
  Review of Materials Research} \textbf{2015}, \emph{45}, 195--216\relax
\mciteBstWouldAddEndPuncttrue
\mciteSetBstMidEndSepPunct{\mcitedefaultmidpunct}
{\mcitedefaultendpunct}{\mcitedefaultseppunct}\relax
\EndOfBibitem
\bibitem[Ghiandoni \latin{et~al.}(2019)Ghiandoni, Bodkin, Chen, Hristozov,
  Wallace, Webster, and Gillet]{ghiandoni2019development}
Ghiandoni,~G.~M.; Bodkin,~M.~J.; Chen,~B.; Hristozov,~D.; Wallace,~J.~E.;
  Webster,~J.; Gillet,~V.~J. Development and application of a data-driven
  reaction classification model: comparison of an electronic lab notebook and
  medicinal chemistry literature. \emph{Journal of chemical information and
  modeling} \textbf{2019}, \emph{59}, 4167--4187\relax
\mciteBstWouldAddEndPuncttrue
\mciteSetBstMidEndSepPunct{\mcitedefaultmidpunct}
{\mcitedefaultendpunct}{\mcitedefaultseppunct}\relax
\EndOfBibitem
\bibitem[Rupp \latin{et~al.}(2012)Rupp, Tkatchenko, M{\"u}ller, and
  Von~Lilienfeld]{Rupp2012}
Rupp,~M.; Tkatchenko,~A.; M{\"u}ller,~K.-R.; Von~Lilienfeld,~O.~A. Fast and
  accurate modeling of molecular atomization energies with machine learning.
  \emph{Physical review letters} \textbf{2012}, \emph{108}, 058301\relax
\mciteBstWouldAddEndPuncttrue
\mciteSetBstMidEndSepPunct{\mcitedefaultmidpunct}
{\mcitedefaultendpunct}{\mcitedefaultseppunct}\relax
\EndOfBibitem
\bibitem[Tsubaki and Mizoguchi(2018)Tsubaki, and Mizoguchi]{Tsukabi2018}
Tsubaki,~M.; Mizoguchi,~T. Fast and accurate molecular property prediction:
  learning atomic interactions and potentials with neural networks. \emph{The
  journal of physical chemistry letters} \textbf{2018}, \emph{9},
  5733--5741\relax
\mciteBstWouldAddEndPuncttrue
\mciteSetBstMidEndSepPunct{\mcitedefaultmidpunct}
{\mcitedefaultendpunct}{\mcitedefaultseppunct}\relax
\EndOfBibitem
\bibitem[Kuzminykh \latin{et~al.}(2018)Kuzminykh, Polykovskiy, Kadurin,
  Zhebrak, Baskov, Nikolenko, Shayakhmetov, and Zhavoronkov]{Kuzminykh20183d}
Kuzminykh,~D.; Polykovskiy,~D.; Kadurin,~A.; Zhebrak,~A.; Baskov,~I.;
  Nikolenko,~S.; Shayakhmetov,~R.; Zhavoronkov,~A. 3d molecular representations
  based on the wave transform for convolutional neural networks.
  \emph{Molecular pharmaceutics} \textbf{2018}, \emph{15}, 4378--4385\relax
\mciteBstWouldAddEndPuncttrue
\mciteSetBstMidEndSepPunct{\mcitedefaultmidpunct}
{\mcitedefaultendpunct}{\mcitedefaultseppunct}\relax
\EndOfBibitem
\bibitem[Wang \latin{et~al.}(2021)Wang, Kauwe, Murdock, and
  Sparks]{wang2021compositionally_crabnet}
Wang,~A. Y.-T.; Kauwe,~S.~K.; Murdock,~R.~J.; Sparks,~T.~D. Compositionally
  restricted attention-based network for materials property predictions.
  \emph{Npj Computational Materials} \textbf{2021}, \emph{7}, 77\relax
\mciteBstWouldAddEndPuncttrue
\mciteSetBstMidEndSepPunct{\mcitedefaultmidpunct}
{\mcitedefaultendpunct}{\mcitedefaultseppunct}\relax
\EndOfBibitem
\bibitem[Chen \latin{et~al.}(2019)Chen, Ye, Zuo, Zheng, and Ong]{Chen2019}
Chen,~C.; Ye,~W.; Zuo,~Y.; Zheng,~C.; Ong,~S.~P. Graph networks as a universal
  machine learning framework for molecules and crystals. \emph{Chemistry of
  Materials} \textbf{2019}, \emph{31}, 3564--3572\relax
\mciteBstWouldAddEndPuncttrue
\mciteSetBstMidEndSepPunct{\mcitedefaultmidpunct}
{\mcitedefaultendpunct}{\mcitedefaultseppunct}\relax
\EndOfBibitem
\bibitem[De~Breuck \latin{et~al.}(2021)De~Breuck, Hautier, and
  Rignanese]{debreuck2021materials}
De~Breuck,~P.-P.; Hautier,~G.; Rignanese,~G.-M. Materials property prediction
  for limited datasets enabled by feature selection and joint learning with
  MODNet. \emph{npj Computational Materials} \textbf{2021}, \emph{7}, 83\relax
\mciteBstWouldAddEndPuncttrue
\mciteSetBstMidEndSepPunct{\mcitedefaultmidpunct}
{\mcitedefaultendpunct}{\mcitedefaultseppunct}\relax
\EndOfBibitem
\bibitem[Maurer \latin{et~al.}(2019)Maurer, Freysoldt, Reilly, Brandenburg,
  Hofmann, Björkman, Lebègue, and Tkatchenko]{Maurer2019}
Maurer,~R.~J.; Freysoldt,~C.; Reilly,~A.~M.; Brandenburg,~J.~G.;
  Hofmann,~O.~T.; Björkman,~T.; Lebègue,~S.; Tkatchenko,~A. Advances in
  Density-Functional Calculations for Materials Modeling. \emph{Annual Review
  of Materials Research} \textbf{2019}, \emph{49}, 1--30\relax
\mciteBstWouldAddEndPuncttrue
\mciteSetBstMidEndSepPunct{\mcitedefaultmidpunct}
{\mcitedefaultendpunct}{\mcitedefaultseppunct}\relax
\EndOfBibitem
\bibitem[Perdew and Levy(1983)Perdew, and Levy]{perdew1983physical}
Perdew,~J.~P.; Levy,~M. Physical content of the exact Kohn-Sham orbital
  energies: band gaps and derivative discontinuities. \emph{Physical Review
  Letters} \textbf{1983}, \emph{51}, 1884\relax
\mciteBstWouldAddEndPuncttrue
\mciteSetBstMidEndSepPunct{\mcitedefaultmidpunct}
{\mcitedefaultendpunct}{\mcitedefaultseppunct}\relax
\EndOfBibitem
\bibitem[Hautier \latin{et~al.}(2012)Hautier, Ong, Jain, Moore, and
  Ceder]{hautier2012accuracy_bartel15}
Hautier,~G.; Ong,~S.~P.; Jain,~A.; Moore,~C.~J.; Ceder,~G. Accuracy of density
  functional theory in predicting formation energies of ternary oxides from
  binary oxides and its implication on phase stability. \emph{Physical Review
  B} \textbf{2012}, \emph{85}, 155208\relax
\mciteBstWouldAddEndPuncttrue
\mciteSetBstMidEndSepPunct{\mcitedefaultmidpunct}
{\mcitedefaultendpunct}{\mcitedefaultseppunct}\relax
\EndOfBibitem
\bibitem[Bartel \latin{et~al.}(2019)Bartel, Weimer, Lany, Musgrave, and
  Holder]{bartel2019therole_bartel16}
Bartel,~C.~J.; Weimer,~A.~W.; Lany,~S.; Musgrave,~C.~B.; Holder,~A.~M. The role
  of decomposition reactions in assessing first-principles predictions of solid
  stability. \emph{npj Computational Materials} \textbf{2019}, \emph{5},
  4\relax
\mciteBstWouldAddEndPuncttrue
\mciteSetBstMidEndSepPunct{\mcitedefaultmidpunct}
{\mcitedefaultendpunct}{\mcitedefaultseppunct}\relax
\EndOfBibitem
\bibitem[Bartel \latin{et~al.}(2020)Bartel, Trewartha, Wang, Dunn, Jain, and
  Ceder]{Bartel2020}
Bartel,~C.~J.; Trewartha,~A.; Wang,~Q.; Dunn,~A.; Jain,~A.; Ceder,~G. A
  critical examination of compound stability predictions from machine-learned
  formation energies. \emph{npj Computational Materials} \textbf{2020},
  \emph{6}, 97\relax
\mciteBstWouldAddEndPuncttrue
\mciteSetBstMidEndSepPunct{\mcitedefaultmidpunct}
{\mcitedefaultendpunct}{\mcitedefaultseppunct}\relax
\EndOfBibitem
\bibitem[Morales-Garc{\'\i}a \latin{et~al.}(2017)Morales-Garc{\'\i}a, Valero,
  and Illas]{morales2017}
Morales-Garc{\'\i}a,~{\'A}.; Valero,~R.; Illas,~F. An empirical, yet practical
  way to predict the band gap in solids by using density functional band
  structure calculations. \emph{The Journal of Physical Chemistry C}
  \textbf{2017}, \emph{121}, 18862--18866\relax
\mciteBstWouldAddEndPuncttrue
\mciteSetBstMidEndSepPunct{\mcitedefaultmidpunct}
{\mcitedefaultendpunct}{\mcitedefaultseppunct}\relax
\EndOfBibitem
\bibitem[Geman \latin{et~al.}(1992)Geman, Bienenstock, and
  Doursat]{geman1992neural}
Geman,~S.; Bienenstock,~E.; Doursat,~R. Neural networks and the bias/variance
  dilemma. \emph{Neural computation} \textbf{1992}, \emph{4}, 1--58\relax
\mciteBstWouldAddEndPuncttrue
\mciteSetBstMidEndSepPunct{\mcitedefaultmidpunct}
{\mcitedefaultendpunct}{\mcitedefaultseppunct}\relax
\EndOfBibitem
\bibitem[Greenman \latin{et~al.}(2022)Greenman, Green, and
  Gomez-Bombarelli]{greenman2022multi_mf1}
Greenman,~K.~P.; Green,~W.~H.; Gomez-Bombarelli,~R. Multi-fidelity prediction
  of molecular optical peaks with deep learning. \emph{Chemical Science}
  \textbf{2022}, \emph{13}, 1152--1162\relax
\mciteBstWouldAddEndPuncttrue
\mciteSetBstMidEndSepPunct{\mcitedefaultmidpunct}
{\mcitedefaultendpunct}{\mcitedefaultseppunct}\relax
\EndOfBibitem
\bibitem[Batra \latin{et~al.}(2019)Batra, Pilania, Uberuaga, and
  Ramprasad]{batra2019multifidelity_mf2}
Batra,~R.; Pilania,~G.; Uberuaga,~B.~P.; Ramprasad,~R. Multifidelity
  information fusion with machine learning: A case study of dopant formation
  energies in hafnia. \emph{ACS applied materials \& interfaces} \textbf{2019},
  \emph{11}, 24906--24918\relax
\mciteBstWouldAddEndPuncttrue
\mciteSetBstMidEndSepPunct{\mcitedefaultmidpunct}
{\mcitedefaultendpunct}{\mcitedefaultseppunct}\relax
\EndOfBibitem
\bibitem[Egorova \latin{et~al.}(2020)Egorova, Hafizi, Woods, and
  Day]{egorova2020multifidelity_mf3}
Egorova,~O.; Hafizi,~R.; Woods,~D.~C.; Day,~G.~M. Multifidelity statistical
  machine learning for molecular crystal structure prediction. \emph{The
  Journal of Physical Chemistry A} \textbf{2020}, \emph{124}, 8065--8078\relax
\mciteBstWouldAddEndPuncttrue
\mciteSetBstMidEndSepPunct{\mcitedefaultmidpunct}
{\mcitedefaultendpunct}{\mcitedefaultseppunct}\relax
\EndOfBibitem
\bibitem[Chen \latin{et~al.}(2021)Chen, Zuo, Ye, Li, and Ong]{chen2021learning}
Chen,~C.; Zuo,~Y.; Ye,~W.; Li,~X.; Ong,~S.~P. Learning properties of ordered
  and disordered materials from multi-fidelity data. \emph{Nature Computational
  Science} \textbf{2021}, \emph{1}, 46--53\relax
\mciteBstWouldAddEndPuncttrue
\mciteSetBstMidEndSepPunct{\mcitedefaultmidpunct}
{\mcitedefaultendpunct}{\mcitedefaultseppunct}\relax
\EndOfBibitem
\bibitem[Tran \latin{et~al.}(2020)Tran, Tranchida, Wildey, and
  Thompson]{tran2020multi_mf4}
Tran,~A.; Tranchida,~J.; Wildey,~T.; Thompson,~A.~P. Multi-fidelity
  machine-learning with uncertainty quantification and Bayesian optimization
  for materials design: Application to ternary random alloys. \emph{The Journal
  of Chemical Physics} \textbf{2020}, \emph{153}, 074705\relax
\mciteBstWouldAddEndPuncttrue
\mciteSetBstMidEndSepPunct{\mcitedefaultmidpunct}
{\mcitedefaultendpunct}{\mcitedefaultseppunct}\relax
\EndOfBibitem
\bibitem[Huang \latin{et~al.}(2019)Huang, Qu, Jia, and
  Zhao]{Huang_2019_ICCV_o2u}
Huang,~J.; Qu,~L.; Jia,~R.; Zhao,~B. O2u-net: A simple noisy label detection
  approach for deep neural networks. Proceedings of the IEEE/CVF International
  Conference on Computer Vision. 2019; pp 3326--3334\relax
\mciteBstWouldAddEndPuncttrue
\mciteSetBstMidEndSepPunct{\mcitedefaultmidpunct}
{\mcitedefaultendpunct}{\mcitedefaultseppunct}\relax
\EndOfBibitem
\bibitem[Oja(1980)]{erkki1980_first_noise}
Oja,~E. On the convergence of an associative learning algorithm in the presence
  of noise. \emph{International Journal of Systems Science} \textbf{1980},
  \emph{11}, 629--640\relax
\mciteBstWouldAddEndPuncttrue
\mciteSetBstMidEndSepPunct{\mcitedefaultmidpunct}
{\mcitedefaultendpunct}{\mcitedefaultseppunct}\relax
\EndOfBibitem
\bibitem[Angluin and Laird(1988)Angluin, and Laird]{angluin1988}
Angluin,~D.; Laird,~P. Learning From Noisy Examples. \emph{Machine Learning}
  \textbf{1988}, \emph{2}, 343--370\relax
\mciteBstWouldAddEndPuncttrue
\mciteSetBstMidEndSepPunct{\mcitedefaultmidpunct}
{\mcitedefaultendpunct}{\mcitedefaultseppunct}\relax
\EndOfBibitem
\bibitem[Han \latin{et~al.}(2018)Han, Yao, Yu, Niu, Xu, Hu, Tsang, and
  Sugiyama]{han2018coteaching_o2u5}
Han,~B.; Yao,~Q.; Yu,~X.; Niu,~G.; Xu,~M.; Hu,~W.; Tsang,~I.~W.; Sugiyama,~M.
  Co-teaching: robust training of deep neural networks with extremely noisy
  labels. Proceedings of the 32nd International Conference on Neural
  Information Processing Systems. 2018; pp 8536--8546\relax
\mciteBstWouldAddEndPuncttrue
\mciteSetBstMidEndSepPunct{\mcitedefaultmidpunct}
{\mcitedefaultendpunct}{\mcitedefaultseppunct}\relax
\EndOfBibitem
\bibitem[Bengio \latin{et~al.}(2009)Bengio, Louradour, Collobert, and
  Weston]{Bengio2009}
Bengio,~Y.; Louradour,~J.; Collobert,~R.; Weston,~J. Curriculum learning.
  Proceedings of the 26th Annual International Conference on Machine Learning.
  2009; pp 41--48\relax
\mciteBstWouldAddEndPuncttrue
\mciteSetBstMidEndSepPunct{\mcitedefaultmidpunct}
{\mcitedefaultendpunct}{\mcitedefaultseppunct}\relax
\EndOfBibitem
\bibitem[Guo \latin{et~al.}(2018)Guo, Huang, Zhang, Zhuang, Dong, Scott, and
  Huang]{sheng2018curriculumnet_o2u4}
Guo,~S.; Huang,~W.; Zhang,~H.; Zhuang,~C.; Dong,~D.; Scott,~M.~R.; Huang,~D.
  Curriculumnet: Weakly supervised learning from large-scale web images.
  Proceedings of the European Conference on Computer Vision (ECCV). 2018; pp
  135--150\relax
\mciteBstWouldAddEndPuncttrue
\mciteSetBstMidEndSepPunct{\mcitedefaultmidpunct}
{\mcitedefaultendpunct}{\mcitedefaultseppunct}\relax
\EndOfBibitem
\bibitem[Jiang \latin{et~al.}(2018)Jiang, Zhou, Leung, Li, and
  Fei-Fei]{jiang2017mentornet_o2u7}
Jiang,~L.; Zhou,~Z.; Leung,~T.; Li,~L.-J.; Fei-Fei,~L. Mentornet: Learning
  data-driven curriculum for very deep neural networks on corrupted labels.
  International Conference on Machine Learning. 2018; pp 2304--2313\relax
\mciteBstWouldAddEndPuncttrue
\mciteSetBstMidEndSepPunct{\mcitedefaultmidpunct}
{\mcitedefaultendpunct}{\mcitedefaultseppunct}\relax
\EndOfBibitem
\bibitem[Donoho(1995)]{donoho1995noising}
Donoho,~D.~L. De-noising by soft-thresholding. \emph{IEEE transactions on
  information theory} \textbf{1995}, \emph{41}, 613--627\relax
\mciteBstWouldAddEndPuncttrue
\mciteSetBstMidEndSepPunct{\mcitedefaultmidpunct}
{\mcitedefaultendpunct}{\mcitedefaultseppunct}\relax
\EndOfBibitem
\bibitem[Donoho and Johnstone(1994)Donoho, and Johnstone]{donoho1994ideal}
Donoho,~D.~L.; Johnstone,~J.~M. Ideal spatial adaptation by wavelet shrinkage.
  \emph{biometrika} \textbf{1994}, \emph{81}, 425--455\relax
\mciteBstWouldAddEndPuncttrue
\mciteSetBstMidEndSepPunct{\mcitedefaultmidpunct}
{\mcitedefaultendpunct}{\mcitedefaultseppunct}\relax
\EndOfBibitem
\bibitem[sur()]{sure_example}
A good example can be found at: \url{https://github.com/ilkerbayram/SURE}\relax
\mciteBstWouldAddEndPuncttrue
\mciteSetBstMidEndSepPunct{\mcitedefaultmidpunct}
{\mcitedefaultendpunct}{\mcitedefaultseppunct}\relax
\EndOfBibitem
\bibitem[Perdew \latin{et~al.}(1996)Perdew, Burke, and
  Ernzerhof]{perdew1996generalized_pbe}
Perdew,~J.~P.; Burke,~K.; Ernzerhof,~M. Generalized gradient approximation made
  simple. \emph{Physical review letters} \textbf{1996}, \emph{77}, 3865\relax
\mciteBstWouldAddEndPuncttrue
\mciteSetBstMidEndSepPunct{\mcitedefaultmidpunct}
{\mcitedefaultendpunct}{\mcitedefaultseppunct}\relax
\EndOfBibitem
\bibitem[Heyd \latin{et~al.}(2003)Heyd, Scuseria, and
  Ernzerhof]{heyd2003hybrid_hse}
Heyd,~J.; Scuseria,~G.~E.; Ernzerhof,~M. Hybrid functionals based on a screened
  Coulomb potential. \emph{The Journal of chemical physics} \textbf{2003},
  \emph{118}, 8207--8215\relax
\mciteBstWouldAddEndPuncttrue
\mciteSetBstMidEndSepPunct{\mcitedefaultmidpunct}
{\mcitedefaultendpunct}{\mcitedefaultseppunct}\relax
\EndOfBibitem
\bibitem[jie(2019)]{jie2019_hse1}
A new MaterialGo database and its comparison with other high-throughput
  electronic structure databases for their predicted energy band gaps.
  \emph{Science China Technological Sciences} \textbf{2019}, \emph{62},
  1423--1430\relax
\mciteBstWouldAddEndPuncttrue
\mciteSetBstMidEndSepPunct{\mcitedefaultmidpunct}
{\mcitedefaultendpunct}{\mcitedefaultseppunct}\relax
\EndOfBibitem
\bibitem[Sun \latin{et~al.}(2015)Sun, Ruzsinszky, and
  Perdew]{sun2015strongly_scan}
Sun,~J.; Ruzsinszky,~A.; Perdew,~J.~P. Strongly constrained and appropriately
  normed semilocal density functional. \emph{Physical review letters}
  \textbf{2015}, \emph{115}, 036402\relax
\mciteBstWouldAddEndPuncttrue
\mciteSetBstMidEndSepPunct{\mcitedefaultmidpunct}
{\mcitedefaultendpunct}{\mcitedefaultseppunct}\relax
\EndOfBibitem
\bibitem[Gritsenko \latin{et~al.}(1995)Gritsenko, van Leeuwen, van Lenthe, and
  Baerends]{gritsenko1995self_gllbsc}
Gritsenko,~O.; van Leeuwen,~R.; van Lenthe,~E.; Baerends,~E.~J. Self-consistent
  approximation to the Kohn-Sham exchange potential. \emph{Physical Review A}
  \textbf{1995}, \emph{51}, 1944\relax
\mciteBstWouldAddEndPuncttrue
\mciteSetBstMidEndSepPunct{\mcitedefaultmidpunct}
{\mcitedefaultendpunct}{\mcitedefaultseppunct}\relax
\EndOfBibitem
\bibitem[Kuisma \latin{et~al.}(2010)Kuisma, Ojanen, Enkovaara, and
  Rantala]{kuisma2010kohnsham_gllbsc2}
Kuisma,~M.; Ojanen,~J.; Enkovaara,~J.; Rantala,~T. Kohn-Sham potential with
  discontinuity for band gap materials. \emph{Physical Review B} \textbf{2010},
  \emph{82}, 115106\relax
\mciteBstWouldAddEndPuncttrue
\mciteSetBstMidEndSepPunct{\mcitedefaultmidpunct}
{\mcitedefaultendpunct}{\mcitedefaultseppunct}\relax
\EndOfBibitem
\bibitem[Jain \latin{et~al.}(2013)Jain, Ong, Hautier, Chen, Richards, Dacek,
  Cholia, Gunter, Skinner, Ceder, \latin{et~al.} others]{jain2013commentary_mp}
Jain,~A.; Ong,~S.~P.; Hautier,~G.; Chen,~W.; Richards,~W.~D.; Dacek,~S.;
  Cholia,~S.; Gunter,~D.; Skinner,~D.; Ceder,~G. \latin{et~al.}  Commentary:
  The Materials Project: A materials genome approach to accelerating materials
  innovation. \emph{APL materials} \textbf{2013}, \emph{1}, 011002\relax
\mciteBstWouldAddEndPuncttrue
\mciteSetBstMidEndSepPunct{\mcitedefaultmidpunct}
{\mcitedefaultendpunct}{\mcitedefaultseppunct}\relax
\EndOfBibitem
\bibitem[Borlido \latin{et~al.}(2019)Borlido, Aull, Huran, Tran, Marques, and
  Botti]{borlido2019large_scan1}
Borlido,~P.; Aull,~T.; Huran,~A.~W.; Tran,~F.; Marques,~M.~A.; Botti,~S.
  Large-scale benchmark of exchange--correlation functionals for the
  determination of electronic band gaps of solids. \emph{Journal of chemical
  theory and computation} \textbf{2019}, \emph{15}, 5069--5079\relax
\mciteBstWouldAddEndPuncttrue
\mciteSetBstMidEndSepPunct{\mcitedefaultmidpunct}
{\mcitedefaultendpunct}{\mcitedefaultseppunct}\relax
\EndOfBibitem
\bibitem[Castelli \latin{et~al.}(2015)Castelli, H{\"u}ser, Pandey, Li,
  Thygesen, Seger, Jain, Persson, Ceder, and Jacobsen]{castelli2015new_gllbsc3}
Castelli,~I.~E.; H{\"u}ser,~F.; Pandey,~M.; Li,~H.; Thygesen,~K.~S.; Seger,~B.;
  Jain,~A.; Persson,~K.~A.; Ceder,~G.; Jacobsen,~K.~W. New light-harvesting
  materials using accurate and efficient bandgap calculations. \emph{Advanced
  Energy Materials} \textbf{2015}, \emph{5}, 1400915\relax
\mciteBstWouldAddEndPuncttrue
\mciteSetBstMidEndSepPunct{\mcitedefaultmidpunct}
{\mcitedefaultendpunct}{\mcitedefaultseppunct}\relax
\EndOfBibitem
\bibitem[Zhuo \latin{et~al.}(2018)Zhuo, Mansouri~Tehrani, and
  Brgoch]{zhuo2018predicting}
Zhuo,~Y.; Mansouri~Tehrani,~A.; Brgoch,~J. Predicting the band gaps of
  inorganic solids by machine learning. \emph{The journal of physical chemistry
  letters} \textbf{2018}, \emph{9}, 1668--1673\relax
\mciteBstWouldAddEndPuncttrue
\mciteSetBstMidEndSepPunct{\mcitedefaultmidpunct}
{\mcitedefaultendpunct}{\mcitedefaultseppunct}\relax
\EndOfBibitem
\bibitem[Kingsbury \latin{et~al.}(2022)Kingsbury, Gupta, Bartel, Munro,
  Dwaraknath, Horton, and Persson]{kingsbury2022}
Kingsbury,~R.; Gupta,~A.~S.; Bartel,~C.~J.; Munro,~J.~M.; Dwaraknath,~S.;
  Horton,~M.; Persson,~K.~A. Performance comparison of r2SCAN and SCAN metaGGA
  density functionals for solid materials via an automated, high-throughput
  computational workflow. \emph{Phys. Rev. Mater.} \textbf{2022}, \emph{6},
  013801\relax
\mciteBstWouldAddEndPuncttrue
\mciteSetBstMidEndSepPunct{\mcitedefaultmidpunct}
{\mcitedefaultendpunct}{\mcitedefaultseppunct}\relax
\EndOfBibitem
\bibitem[Lejaeghere \latin{et~al.}(2014)Lejaeghere, Van~Speybroeck, Van~Oost,
  and Cottenier]{lejaeghere2014error}
Lejaeghere,~K.; Van~Speybroeck,~V.; Van~Oost,~G.; Cottenier,~S. Error estimates
  for solid-state density-functional theory predictions: an overview by means
  of the ground-state elemental crystals. \emph{Critical reviews in solid state
  and materials sciences} \textbf{2014}, \emph{39}, 1--24\relax
\mciteBstWouldAddEndPuncttrue
\mciteSetBstMidEndSepPunct{\mcitedefaultmidpunct}
{\mcitedefaultendpunct}{\mcitedefaultseppunct}\relax
\EndOfBibitem
\bibitem[Huerta-Cepas \latin{et~al.}(2016)Huerta-Cepas, Serra, and
  Bork]{huerta2016ete}
Huerta-Cepas,~J.; Serra,~F.; Bork,~P. ETE 3: reconstruction, analysis, and
  visualization of phylogenomic data. \emph{Molecular biology and evolution}
  \textbf{2016}, \emph{33}, 1635--1638\relax
\mciteBstWouldAddEndPuncttrue
\mciteSetBstMidEndSepPunct{\mcitedefaultmidpunct}
{\mcitedefaultendpunct}{\mcitedefaultseppunct}\relax
\EndOfBibitem
\end{mcitethebibliography}

%\end{document}

%\title{A simple denoising approach to exploit multi-fidelity data for machine learning materials properties\\
%(Supporting Information)}

%\begin{document}

%\maketitle

\setcounter{figure}{0}
\setcounter{table}{0}
\setcounter{equation}{0}
\setcounter{section}{0}
\renewcommand{\thefigure}{S\arabic{figure}}
\renewcommand{\thetable}{S\arabic{table}}
\renewcommand{\theequation}{S\arabic{equation}}
\newcommand{\0}{\phantom{0}}

\renewcommand\thesection{\Alph{section}}
\renewcommand\thesubsection{\thesection.\arabic{subsection}}

\section{Supplementary datasets analysis}\label{sec:dataset_analysis}

In this section, we provide various supplementary analyses of the datasets.
These aim both at checking for possible biases in the latter and at delivering more detail about them.

\subsection{Analysis by Element}\label{sec:elements_analysis}

\begin{figure}[H]
\includegraphics{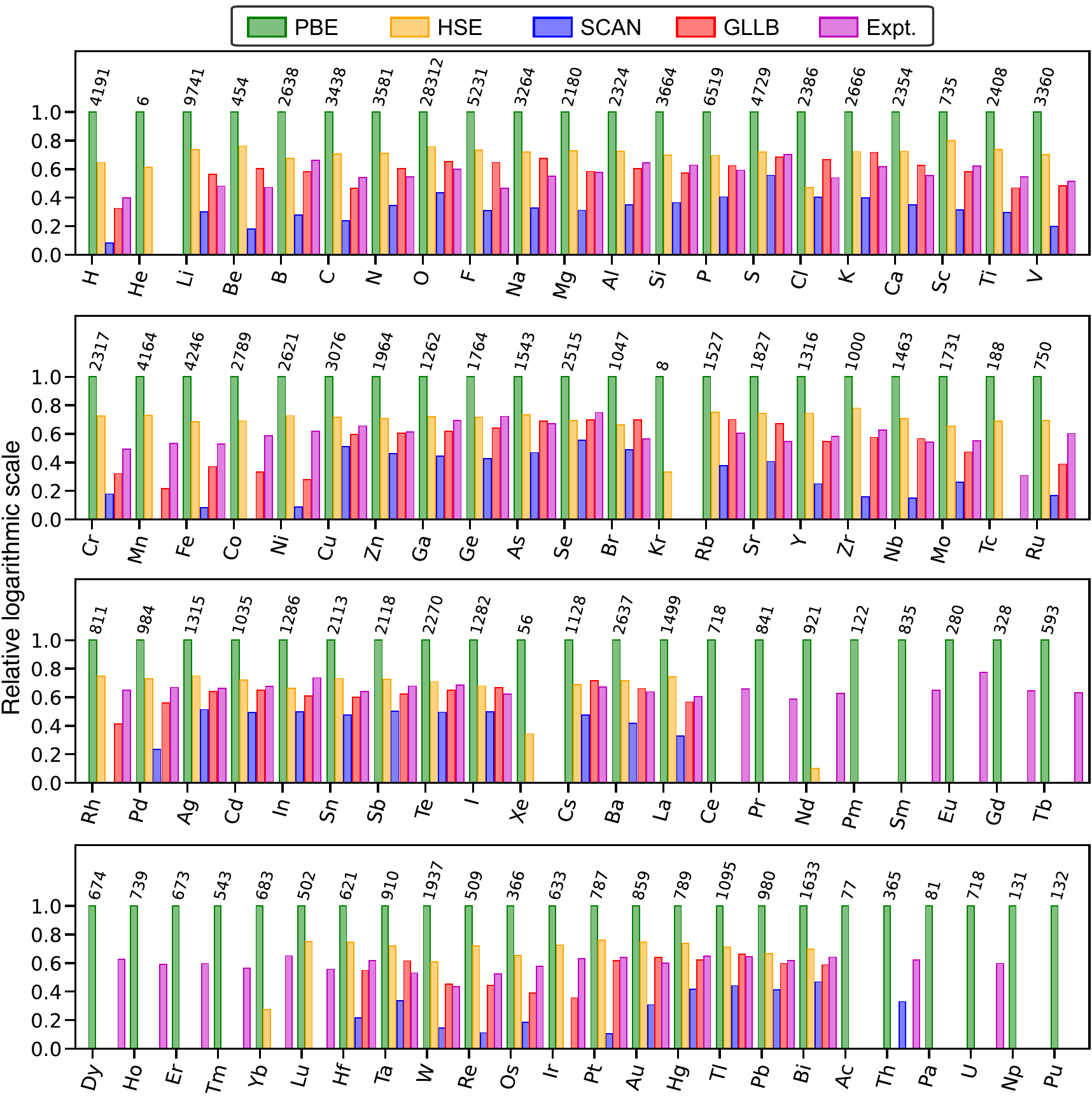}
\caption{Distribution of the chemical elements in the different datasets.
The number over the green bar is the count of structures with that element in the dataset P. For one given element, the height of the bars gives the relative logarithmic ratio of that element in the different datasets.}
\label{fig:ele_distribution}
\end{figure}

The distribution of the chemical elements in each dataset is shown in Fig.~\ref{fig:ele_distribution}. 
Given that the number of structures in the dataset P is considerably larger than in the other datasets, we use it as a reference (normalized to one with the actual number being indicated on top) and we adopt logarithm to indicate the other numbers. 
The PBE dataset contains 87 different chemical elements which cover all the elements present in the other datasets (75 for H, 66 for S, 63 for G, and 80 for E).
All the chemical elements in the dataset E are included in at least one of the DFT datasets ensures that the training procedure and the prediction models of the current work are reasonable.

\subsection{Venn diagram and Upset plot analysis}\label{sec:venn_upset_analysis}

\begin{figure}[H]
\includegraphics[width=0.9\textwidth]{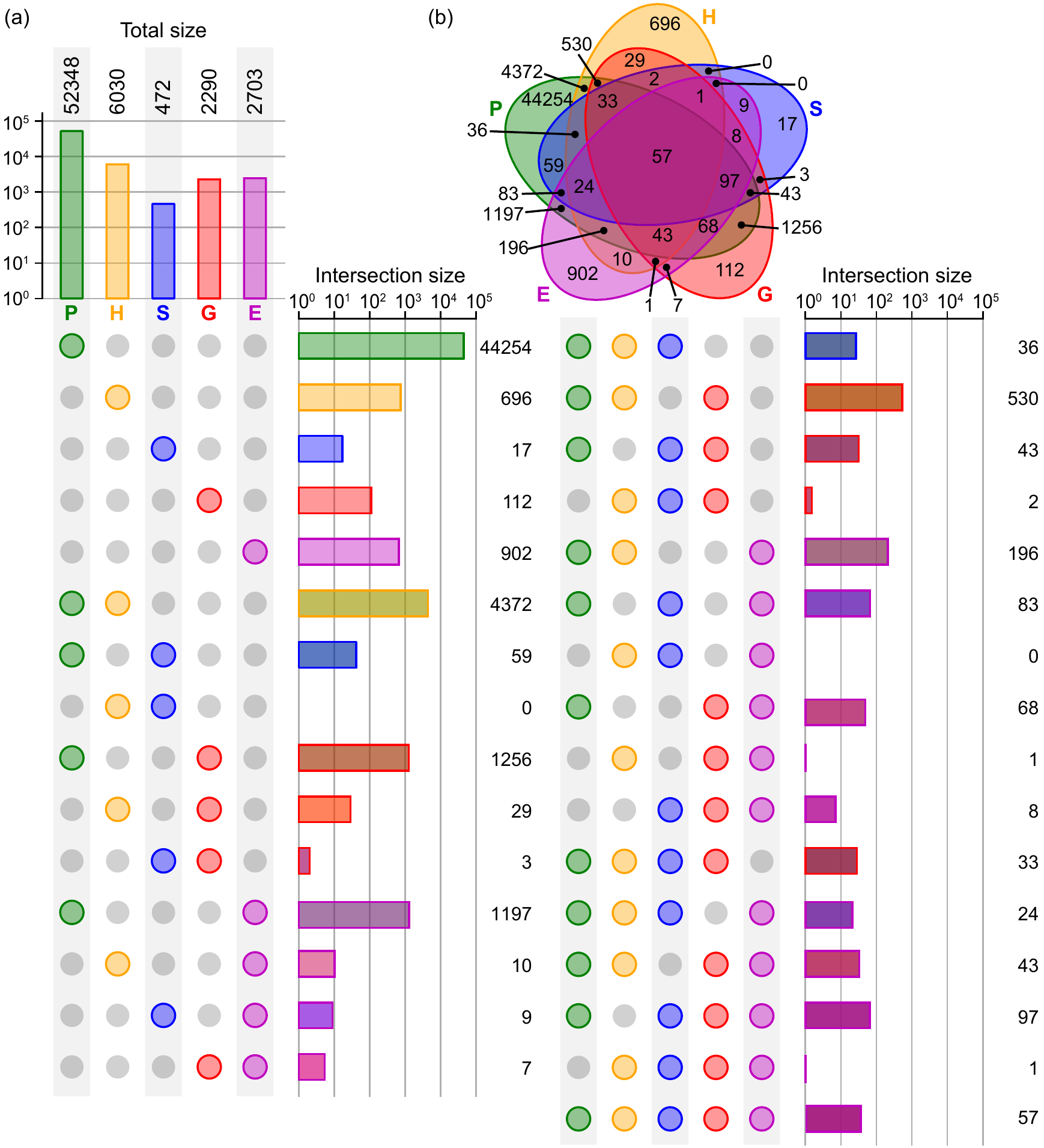}
\caption{(a) Upset plot and (b) Venn diagram illustrating the intersections between the five datasets (P, H, S, G, and the part of E with likely-mpid). All the bar plots are in logarithmic scale. The intersections are illustrated more clearly in Figs.~\ref{fig:venn2}, \ref{fig:venn3}, and \ref{fig:venn4} below.}
\label{fig:venn_upset}
\end{figure}

\begin{figure}[H]
\includegraphics[width=0.9\textwidth]{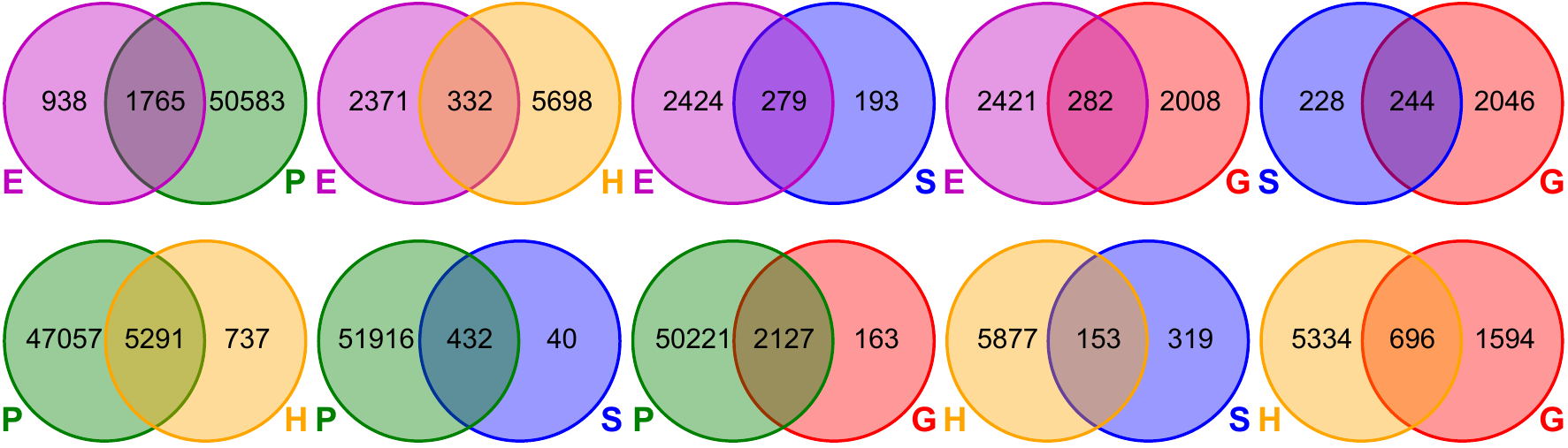}
\caption{The Venn diagrams for any two of the five datasets.}
\label{fig:venn2}
\end{figure}

\begin{figure}[H]
\includegraphics[width=0.9\textwidth]{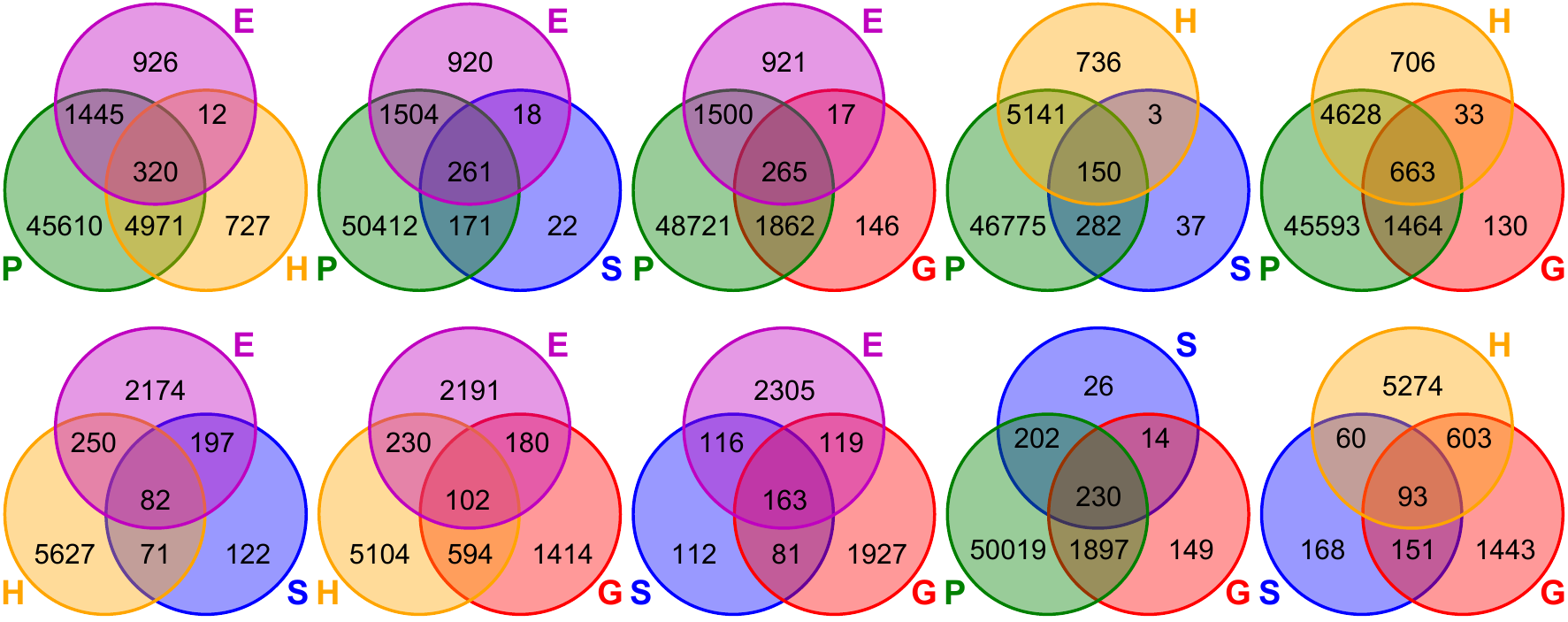}
\caption{The Venn diagrams for any three of the five different datasets.}
\label{fig:venn3}
\end{figure}

\begin{figure}[H]
\includegraphics[width=0.6\textwidth]{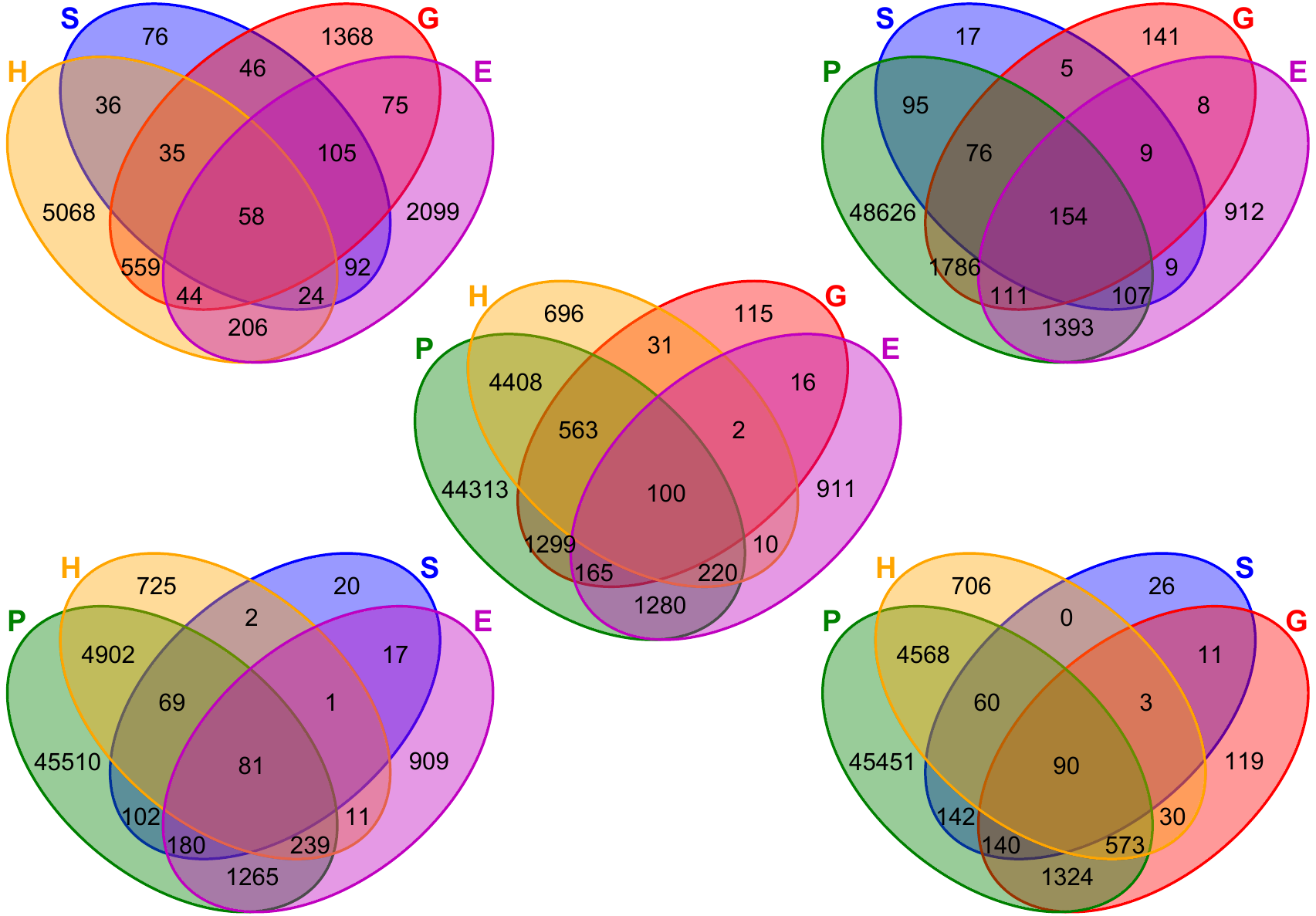}
\caption{The Venn diagrams for any four of the five different datasets.}
\label{fig:venn4}
\end{figure}

To compare the structures available in the different datasets, we first use their Material Project identificators (MP-ids).
For the experimental dataset, 2401 of the 2703 compounds could be assigned a most likely structure from the Materials Project~\cite{kingsbury2022}.
The size of each dataset and of their intersections are shown in Figs.~\ref{fig:venn_upset},  \ref{fig:venn2}, \ref{fig:venn3} and \ref{fig:venn4}.


\subsection{2D distribution of the structures}\label{sec:pca}

\begin{figure}[H]
\includegraphics{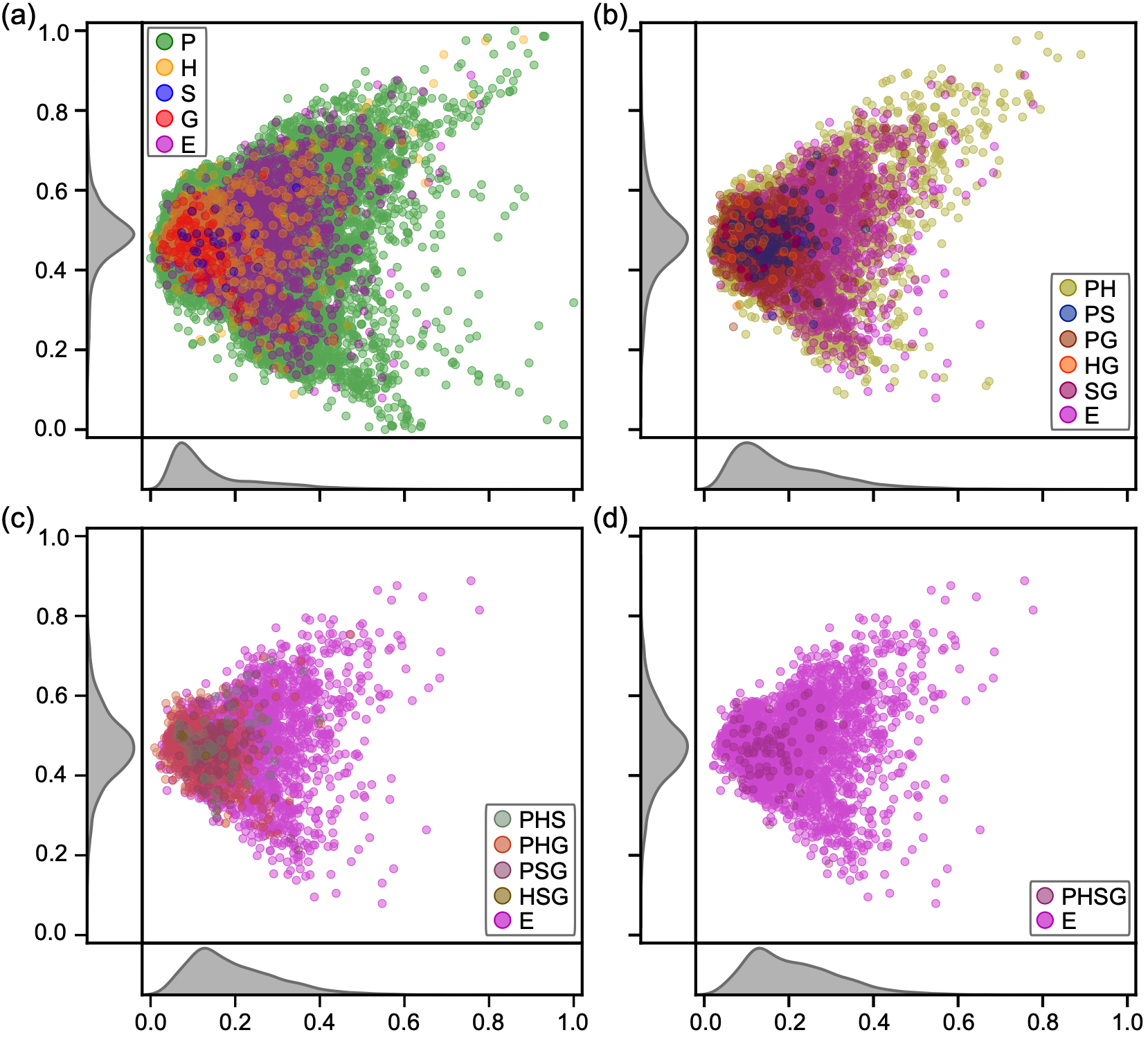}

\caption{Distribution of the datapoints along the 0$^\textrm{th}$ and 1$^\textrm{st}$ PCA directions for the different datasets. E corresponds to the complete experimental dataset in each panel. In panel~(a), P, H, G, and S refer only to the structures that are not included in the other datasets. The panels.~(b), (c), and (d) show the structures that belong to 2, 3, and 4 datasets, respectively. The distributions of the data in each direction are reported in the subplots on the side and at the bottom.}
\label{fig:plt_pca01}
\end{figure}

\begin{figure}[H]
\includegraphics[width=0.8\textwidth]{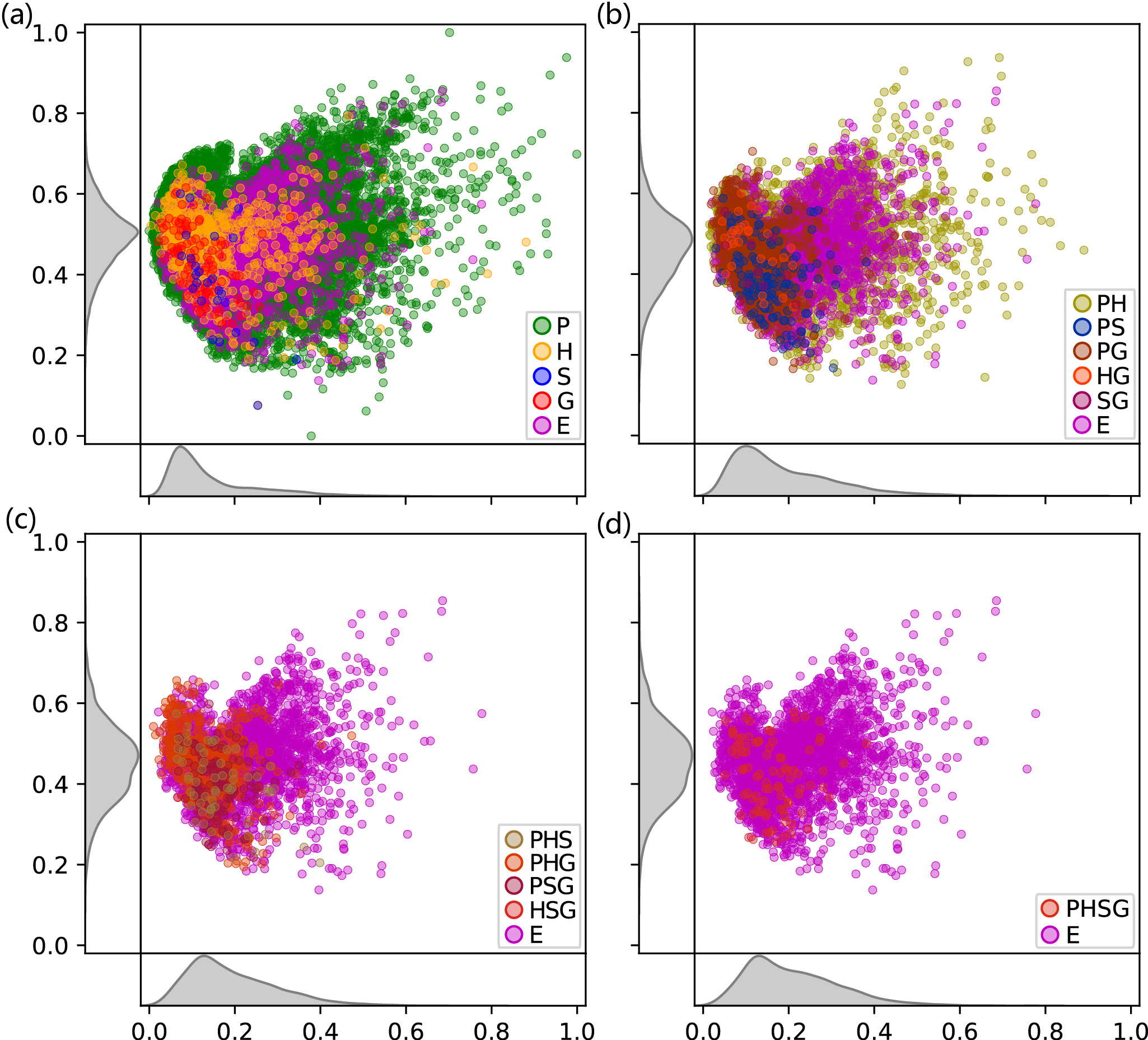}
\caption{Distribution of the datapoints along the 0$^\textrm{th}$ and 2$^\textrm{nd}$ PCA directions for the different datasets. E corresponds to the complete experimental dataset in each panel. In panel~(a), P, H, G, and S refer only to the structures that are not included in the other datasets. The panels.~(b), (c), and (d) show the structures that belong to 2, 3, and 4 datasets, respectively. The distributions of the data in each direction are reported in the subplots on the side and at the bottom.}
\label{fig:plt_pca02}
\end{figure}

\begin{figure}[H]
\includegraphics[width=0.8\textwidth]{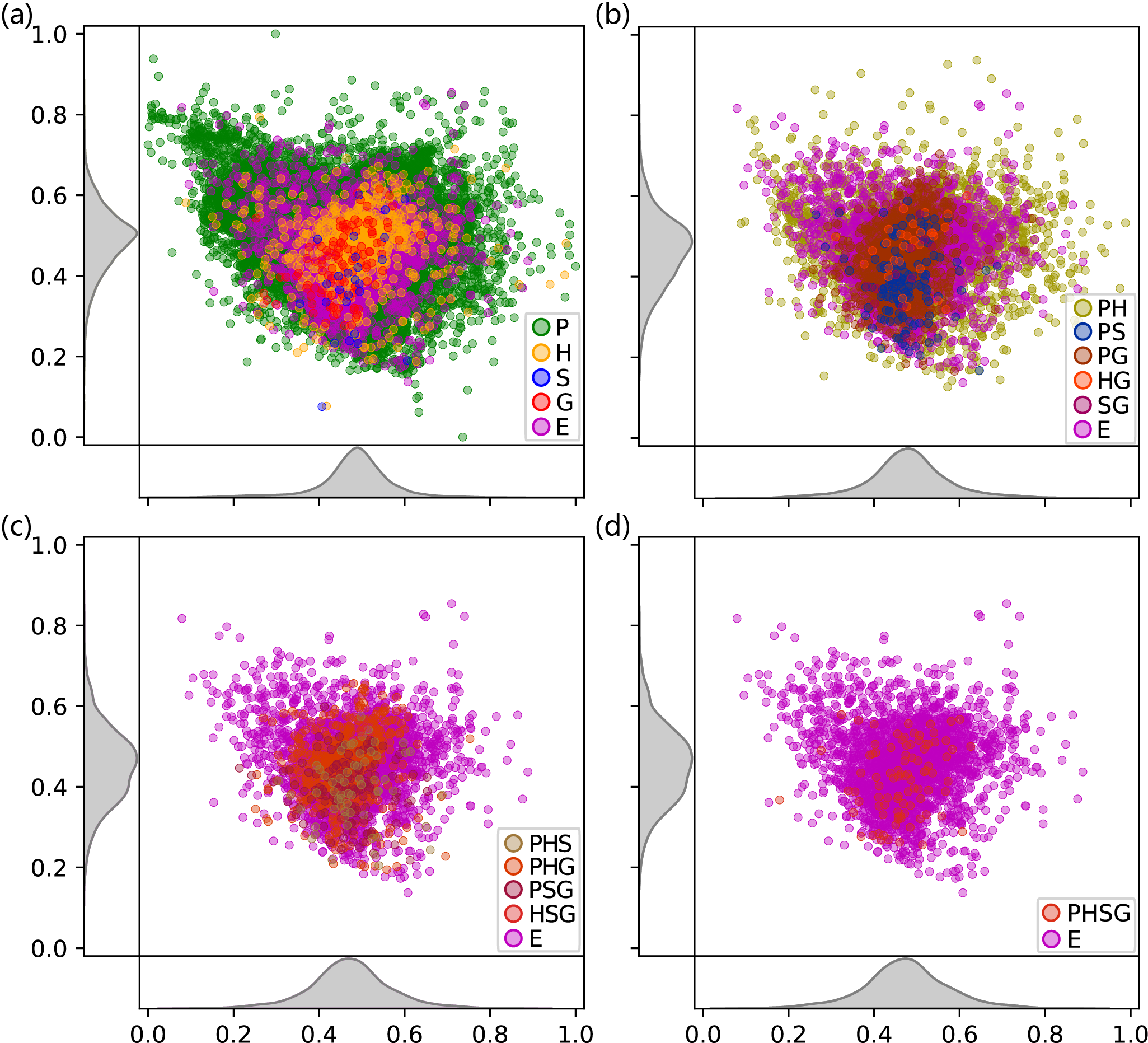}
\caption{Distribution  of the datapoints along the 1$^\textrm{st}$ and 2$^\textrm{nd}$ PCA directions for the different datasets. E corresponds to the complete experimental dataset in each panel. In panel~(a), P, H, G, and S refer only to the structures that are not included in the other datasets. The panels.~(b), (c), and (d) show the structures that belong to 2, 3, and 4 datasets, respectively. The distributions of the data in each direction are reported in the subplots on the side and at the bottom.}
\label{fig:plt_pca12}
\end{figure}

For the structures that have not been assigned an MP-id, it is, however, not possible to figure out how similar it is compared to other structures in the dataset.
To overcome this limitation, we first extract a 96D vector for each structure from a median layer of the MEGNet model. 
Subsequently , we perform a dimensionality reduction through Principal Component Analysis (PCA) and we plot any two of the top three PCA directions. 
The resulting distributions of the data points are shown in Figs.~\ref{fig:plt_pca01}, \ref{fig:plt_pca02} and \ref{fig:plt_pca12}.
Given that, in the PCA approach, the 0$^\textrm{th}$ dimension has the largest variance of all, the mean value in that direction is far from 0.5 after normalization.
The plots provide valuable information about the coverage of the chemical space by all the datasets.
The same trend emerges that P is the most diverse and it covers almost all structures in the other datasets.

\subsection{KL divergence analysis}\label{sec:KL}

Without having to perform dimensionality reduction, the Kullback-Leibler (KL) divergence can also be used to obtain a measure the similarity of two discrete distributions in the previously mentioned 96D space. 
Typically, for two distributions $P$ and $Q$ in the same probability space $\chi$, the KL divergence $D_\mathrm{KL}(P||Q)$ is defined by:
\begin{equation}\label{eq:kl}
    D_\mathrm{KL}(P||Q)=\sum_{x \in \chi} P(x)\ln\frac{P(x)}{Q(x)}.
\end{equation}
Given that $D_\mathrm{KL}(P||Q)$ is not equal to $D_\mathrm{KL}(Q||P)$, we adopt:
\begin{equation}
D(P, Q) = \frac{D_{KL}(P||Q) + D_{KL}(Q||P)}{2}
\end{equation}
as the measure the similarity.
The smaller $D(P, Q)$, the more two distributions are similar with $D(P, P)$=0.
We list some typical $D(P, Q)$ in Table~\ref{tab:KL}.

\begin{table}[h]
    \begin{tabular}{lcccc}
    \hline
                             & \multicolumn{4}{c}{$X$} \\ \cline{2-5}
                             & P     & S     & H     & G     \\
    \hline
        $D(E, X)$            & 0.067 & 0.105 & 0.031 & 0.129 \\
        $D(\overline{X}, X)$ & 0.017 & 0.116 & 0.018 & 0.061 \\
    \hline
    \end{tabular}
    \caption{Bi-directional KL divergence between the distribution of the different datasets, as represented by the MEGNet median layer in 96D space. $X$ refers to one of the DFT datasets (P, S, H and G), while $\overline{X}$ indicates the sum of all the datasets but $X$. $D(E, X)$ is a measure of the similarity between the structures in the experimental and DFT datasets. H with the smallest KL divergence is the most similar to E. $D(\overline{X}, X)$ is a measure of the similarity between structures in a DFT dataset and its complementary set. S with the highest KL divergence is the least similar to all other, and hence brings in the most new information. }
     \label{tab:KL}
\end{table}

\subsection{Distribution of the band gap predictions and errors for the different DFT functionals}\label{sec:dft_distr}

\begin{figure}[!htbp]
\includegraphics[width=1.0\textwidth]{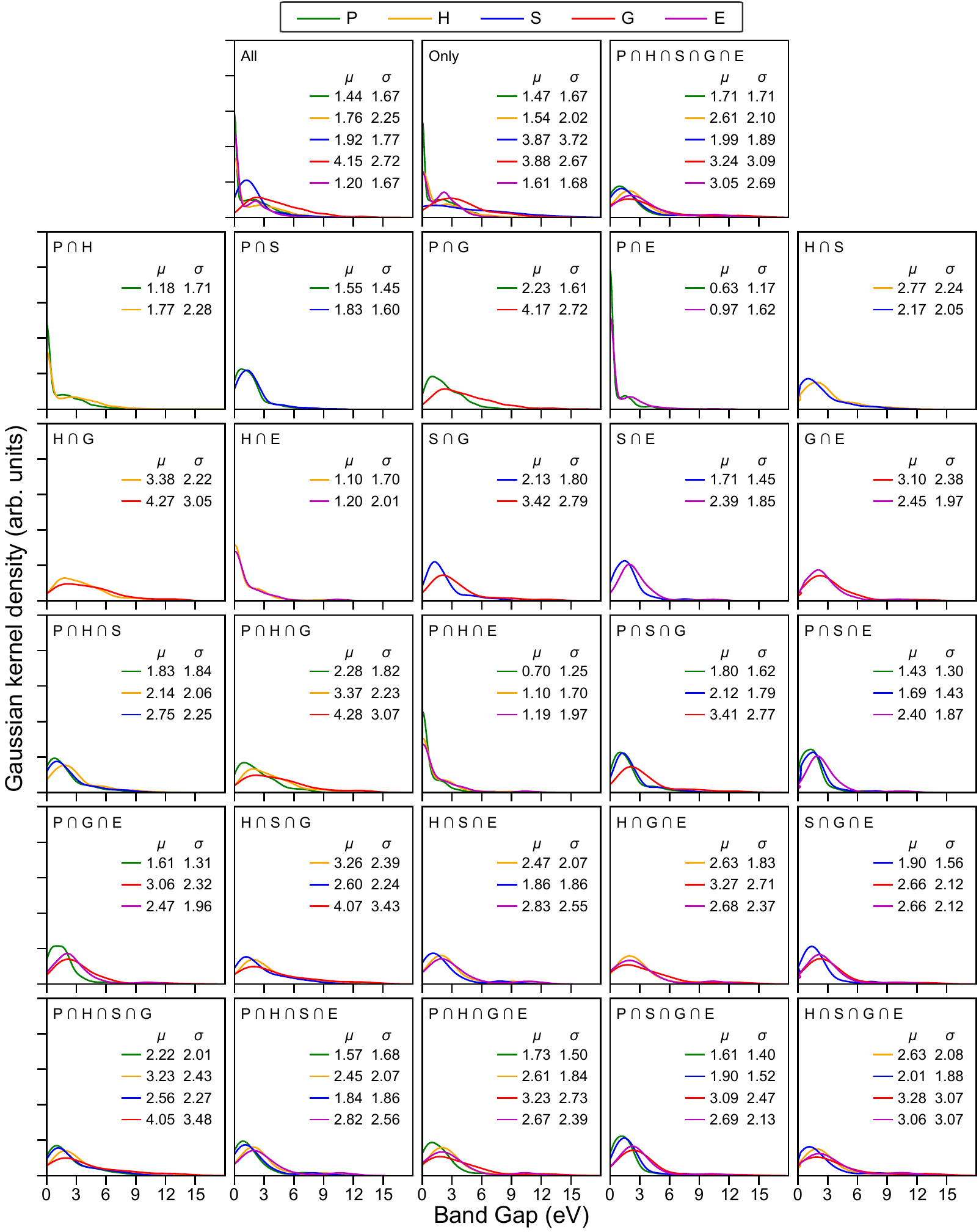}
\caption{An analysis of band gap value distribution with all full and sub datasets in details.}
\label{fig:band_gap_distr}
\end{figure}

\begin{figure}[H]
\includegraphics[width=1.0\textwidth]{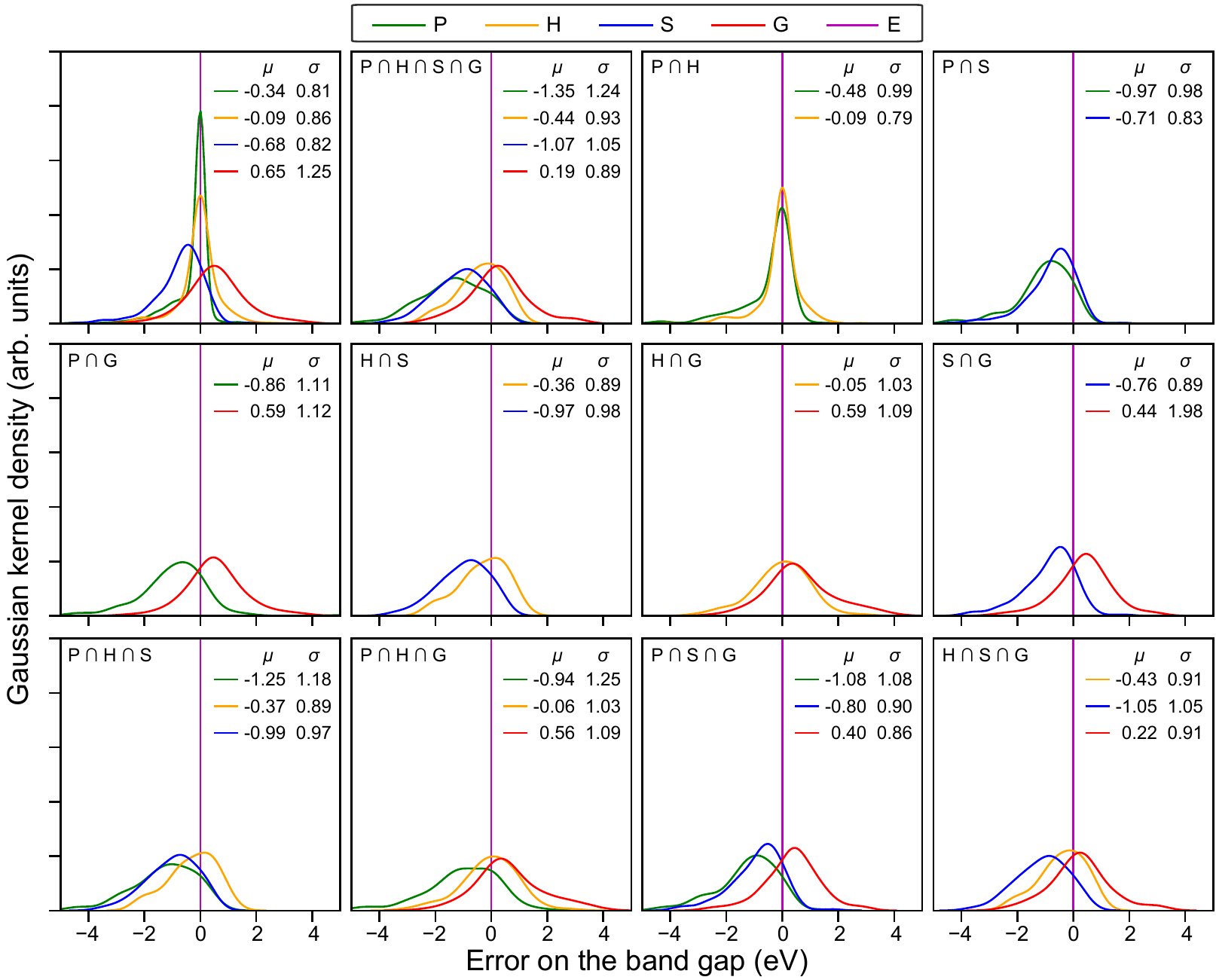}
\caption{An analysis of band gap prediction error (X - E) distribution with all full and sub datasets in details.}
\label{fig:band_gap_err}
\end{figure}

Finally, it is interesting to compare the different DFT predictions for the same structures (even when the experimental data is not available).
An obvious approach to do so is to analyze the distribution of the data points in the different intersections (based on the MP-ids).
% In Table~\ref{tab:intersection_data}, 
In Fig.~\ref{fig:band_gap_distr}, we report the average and standard deviation values for the different intersectionis of the datasets.
As a general trend, we observe that P < S < H < G.
This trend holds for both the average and standard deviation.
It shows the order of absolute value of the band gaps in the different datasets.
A similar representation for the distribution of the errors can be found in Fig.~\ref{fig:band_gap_err}.

\section{Onion tree training results from different denoised data}\label{sec:tree}

\begin{figure}[H]
\includegraphics[width=0.9\textwidth]{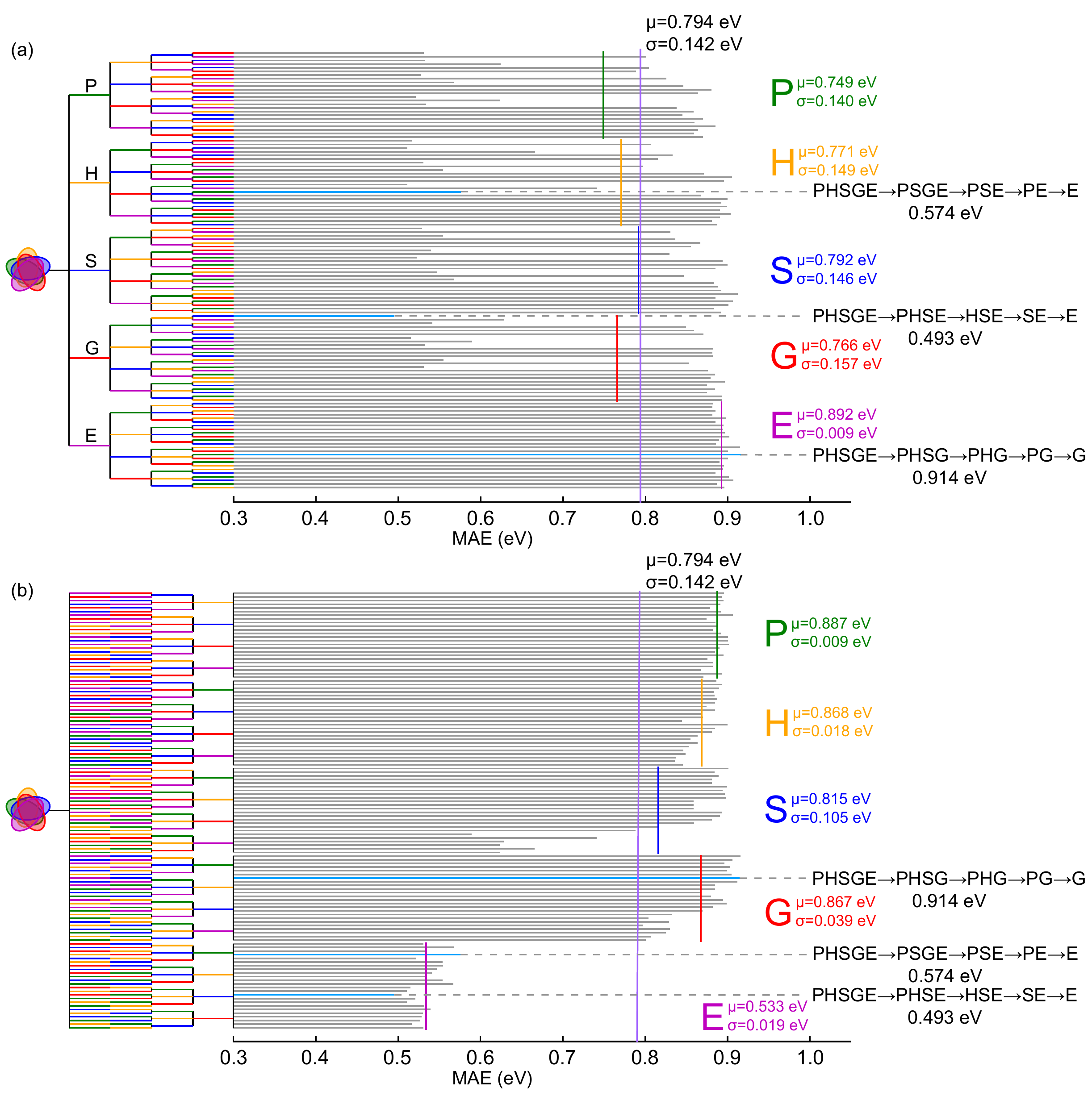}
\caption{MAE results obtained on the data cleaned with the worst model (PHSGE$\rightarrow$PHSG$\rightarrow$PSG$\rightarrow$SG$\rightarrow$G with MAE=0.916~eV) of Fig.~\ref{fig:raw_data_onion_tree_training} using the \textit{onion} training approach for all possible dataset orders: (a) gathered according to the first dataset used and (b) grouped following the last dataset used.
The global average of the MAE is shown by a vertical solid purple line ($\mu$=0.794~eV), while the group averages are indicated by their corresponding color (P in green, H in orange, S in blue, G in red, and E in magenta).
The corresponding standard deviations ($\sigma$) are also indicated accordingly.
The best and worst training sequences, as well as the worst one ending by E, are highlighted in light blue.
The training sequences that produce NaN for one of the folds (so the MAE is only that of the other fold) are indicated by a lighter gray bar, while those that lead to NaN for both folds are left blank.
%Five short dash lines are plotted to express five average values of subtrees. The worst training path is "PHSGE$\rightarrow$PHSG$\rightarrow$PHG$\rightarrow$PG$\rightarrow$G" (MAE is 0.914 eV.), while the best training path is "PHSGE$\rightarrow$PHSE$\rightarrow$HSE$\rightarrow$SE$\rightarrow$E" (MAE is 0.493 eV). Besides previous two path, "PHSGE$\rightarrow$PSGE$\rightarrow$PSE$\rightarrow$PE$\rightarrow$E" (MAE is 0.574 eV) is also highlighted, it is the worst case among all path which ending with dataset 'E'.
}
\label{fig:1st_worst_denoiser_data_tree_training}
\end{figure}

\begin{figure}[H]
\includegraphics[width=0.9\textwidth]{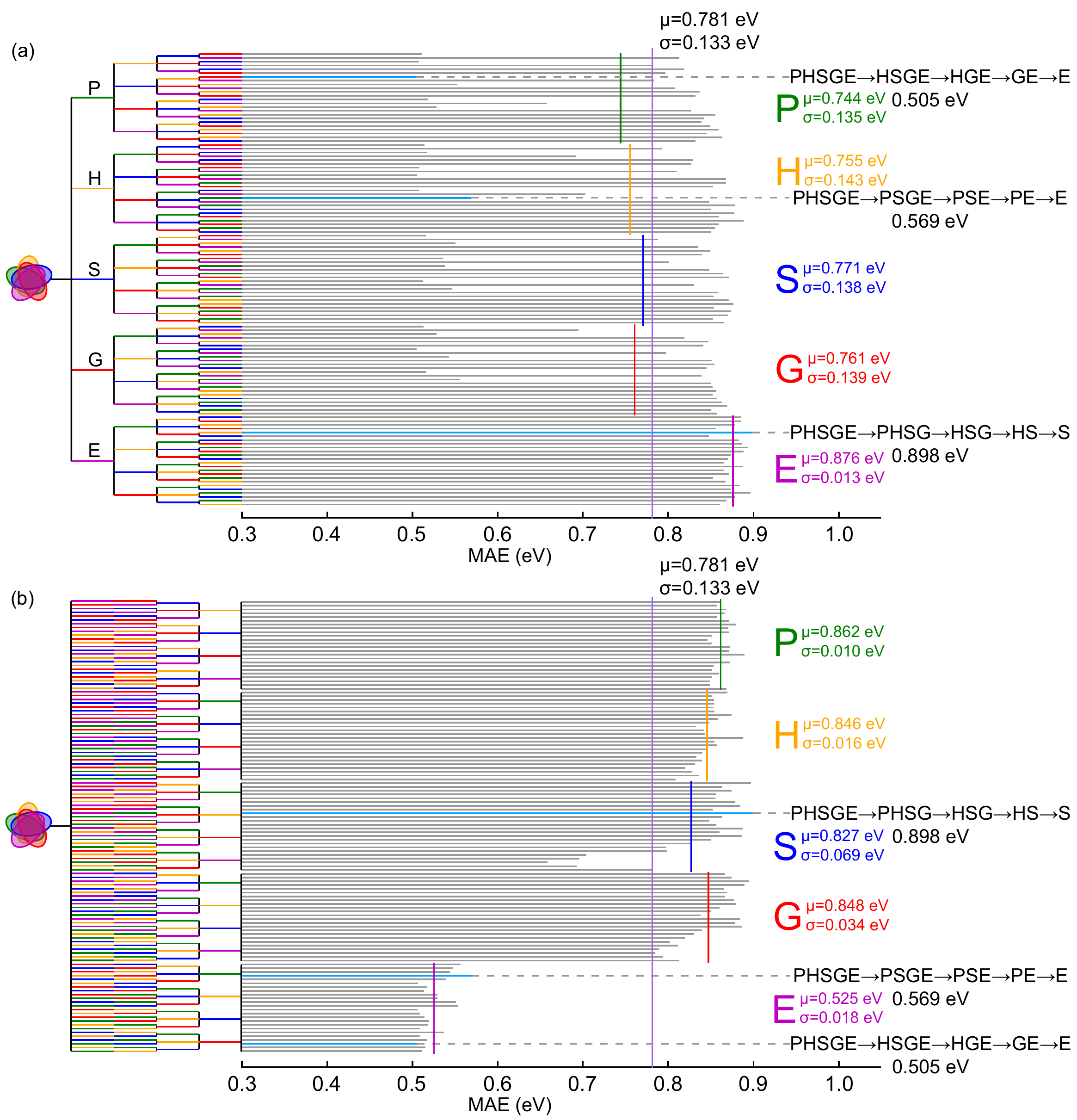}
\caption{MAE results obtained on the data cleaned with the $2^{nd}$ worst model (PHSGE$\rightarrow$PHSG$\rightarrow$PSG$\rightarrow$PG$\rightarrow$G  with MAE=0.889~eV) of Fig.~\ref{fig:raw_data_onion_tree_training} using the \textit{onion} training approach for all possible dataset orders: (a) gathered according to the first dataset used and (b) grouped following the last dataset used.
The global average of the MAE is shown by a vertical solid purple line ($\mu$=0.781~eV), while the group averages are indicated by their corresponding color (P in green, H in orange, S in blue, G in red, and E in magenta).
The corresponding standard deviations ($\sigma$) are also indicated accordingly.
The best and worst training sequences, as well as the worst one ending by E, are highlighted in light blue.
The training sequences that produce NaN for one of the folds (so the MAE is only that of the other fold) are indicated by a lighter gray bar, while those that lead to NaN for both folds are left blank.
%\} of previous 'onion' tree training is used as denoiser.
%The average MAE of all onion training is 0.781 eV (solid vertical line in the figure) comparing with previous average MAE value 0.573 eV without data cleaning (long dash vertical line in the figure). 
%Five short dash lines are plotted to express five average values of subtrees. The worst training path is "PHSGE$\rightarrow$PHSG$\rightarrow$HSG$\rightarrow$HS$\rightarrow$S" (MAE is 0.898 eV.), while the best training path is "PHSGE$\rightarrow$HSGE$\rightarrow$HGE$\rightarrow$GE$\rightarrow$E" (MAE is 0.505 eV). Besides previous two path, "PHSGE$\rightarrow$PSGE$\rightarrow$PSE$\rightarrow$PE$\rightarrow$E" (MAE is 0.569 eV) is also highlighted, it is the worst case among all path which ending with dataset 'E'.
}
\label{fig:2nd_worst_denoiser_data_tree_training}
\end{figure}

\begin{figure}[H]
\includegraphics[width=0.9\textwidth]{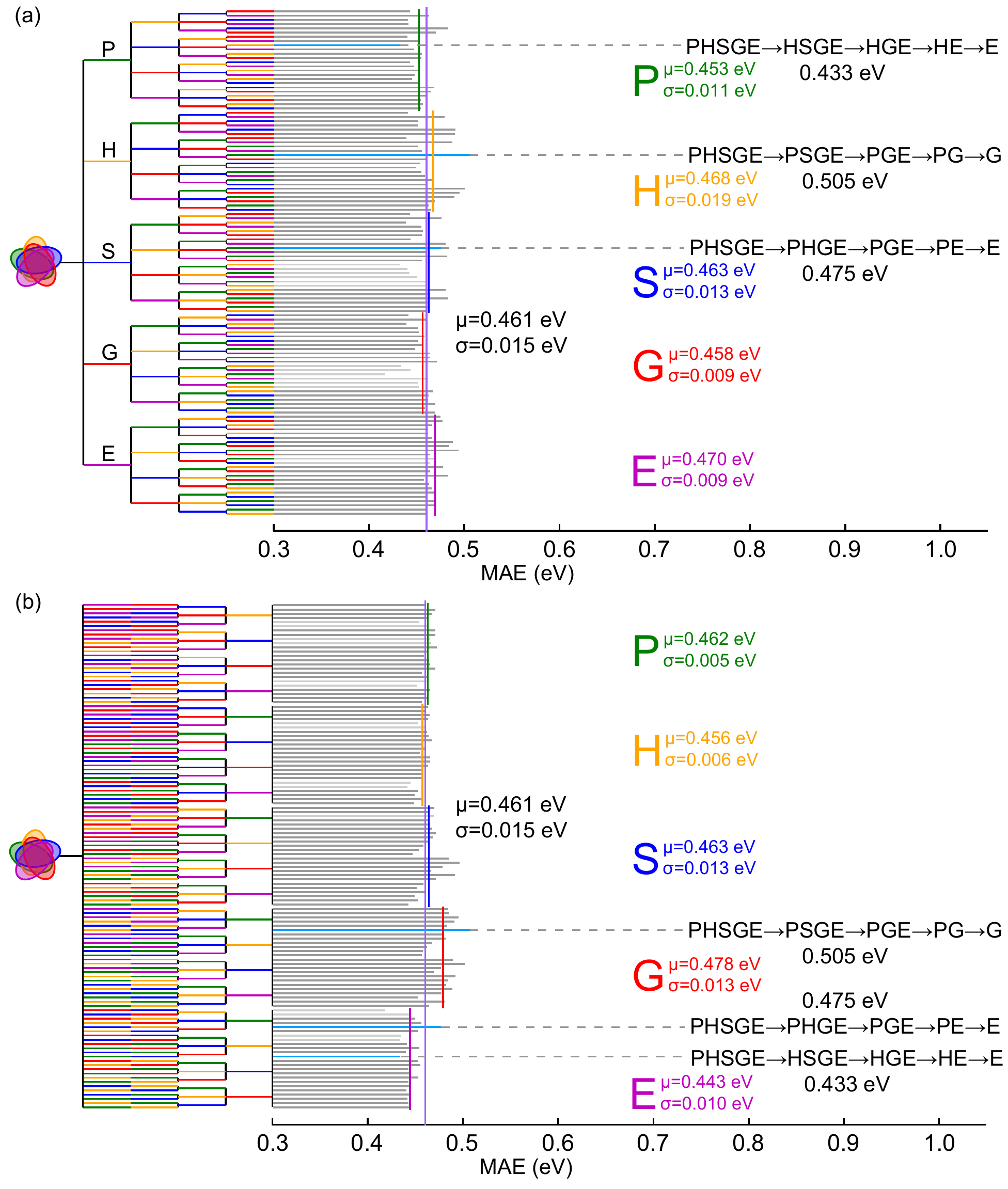}
\caption{MAE results obtained on the data cleaned with a rather poor model among those ending with E (PHSGE$\rightarrow$PHGE$\rightarrow$PGE$\rightarrow$GE$\rightarrow$E with MAE=0.483~eV) of Fig.~\ref{fig:raw_data_onion_tree_training} using the \textit{onion} training approach for all possible dataset orders: (a) gathered according to the first dataset used and (b) grouped following the last dataset used.
The global average of the MAE is shown by a vertical solid purple line ($\mu$=0.461~eV), while the group averages are indicated by their corresponding color (P in green, H in orange, S in blue, G in red, and E in magenta).
The corresponding standard deviations ($\sigma$) are also indicated accordingly.
The best and worst training sequences, as well as the worst one ending by E, are highlighted in light blue.
The training sequences that produce NaN for one of the folds (so the MAE is only that of the other fold) are indicated by a lighter gray bar, while those that lead to NaN for both folds are left blank.
%\} of previous 'onion' tree training is used as denoiser. 
%The average MAE of all onion training is 0.461 eV (solid vertical line in the figure) comparing with previous average MAE value 0.573 eV without data cleaning (long dash vertical line in figure). 
%Five short dash lines are plotted to express five average values of subtrees. The worst training path is "PHSGE$\rightarrow$PSGE$\rightarrow$PGE$\rightarrow$PG$\rightarrow$G" (MAE is 0.505 eV.), while the best training path is "PHSGE$\rightarrow$HSGE$\rightarrow$HGE$\rightarrow$HE$\rightarrow$E" (MAE is 0.433 eV). Besides previous two path, "PHSGE$\rightarrow$PHGE$\rightarrow$PGE$\rightarrow$PE$\rightarrow$E" (MAE is 0.475 eV) is also highlighted, it is the worst case among all path which ending with dataset 'E'.
}
\label{fig:3rd_worst_of_ending_E_denoiser_data_tree_training}
\end{figure}

\begin{figure}[H]
\includegraphics[width=0.9\textwidth]{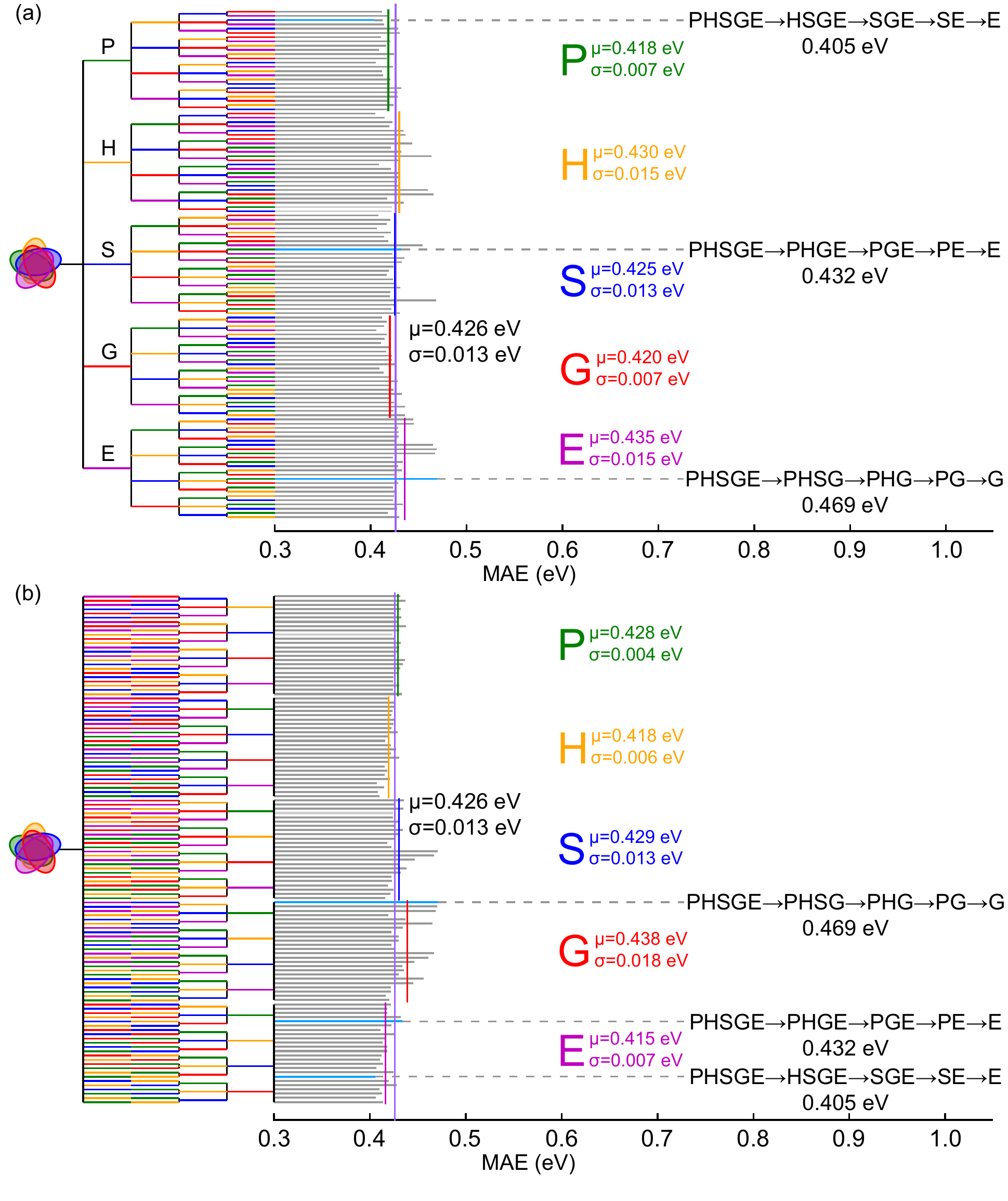}
\caption{MAE results obtained on the data cleaned with a rather good model among those ending with E (PHSGE$\rightarrow$PHSE$\rightarrow$PHE$\rightarrow$HE$\rightarrow$E with MAE=0.443~eV) of Fig.~\ref{fig:raw_data_onion_tree_training} using the \textit{onion} training approach for all possible dataset orders: (a) gathered according to the first dataset used and (b) grouped following the last dataset used.
The global average of the MAE is shown by a vertical solid purple line ($\mu$=0.426~eV), while the group averages are indicated by their corresponding color (P in green, H in orange, S in blue, G in red, and E in magenta).
The corresponding standard deviations ($\sigma$) are also indicated accordingly.
The best and worst training sequences, as well as the worst one ending by E, are highlighted in light blue.
The training sequences that produce NaN for one of the folds (so the MAE is only that of the other fold) are indicated by a lighter gray bar, while those that lead to NaN for both folds are left blank.
%\} of previous 'onion' tree training is used as denoiser. 
%The average MAE of all onion training is 0.426 eV (solid vertical line in the figure) comparing with previous average MAE value 0.573 eV without data cleaning (long dash vertical line in figure). 
%Five short dash lines are plotted to express five average values of subtrees. The worst training path is "PHSGE$\rightarrow$PHSG$\rightarrow$PHG$\rightarrow$PG$\rightarrow$G" (MAE is 0.469 eV), while the best training path is "PHSGE$\rightarrow$HSGE$\rightarrow$SGE$\rightarrow$SE$\rightarrow$E" (MAE is 0.405 eV)
%, an uncompleted route 'EGPHS$\rightarrow$EPHS$\rightarrow$EHS$\rightarrow$EH' reach the lowest MAE in this figure 0.398 eV). 
%Besides previous two path, "PHSGE$\rightarrow$PHGE$\rightarrow$PGE$\rightarrow$PE$\rightarrow$E" (MAE is 0.432 eV) is also highlighted, it is the worst case among all path which ending with dataset 'E'.
}
\label{fig:3rd_best_EGPHS_EPHS_EPH_EH_E_denoiser_onion_tree_training}
\end{figure}

% \newpage
% \bibliography{ref}

\end{document}